**Revealing Symmetry-Broken Superconducting Configurations by Density Functional Theory**

Zi-Kui Liu and Shun-Li Shang*

Department of Materials Science and Engineering, The Pennsylvania State University, University Park, PA 16802, USA
E-mail: sus26@psu.edu

**Abstract:**
A coherent theory for the superconductivity of both conventional and unconventional superconductors is currently lacking. Here we show that superconductivity arises from the formation of a *symmetry-broken* superconducting configuration (SCC) due to atomic perturbation of the normal conducting configuration (NCC). This electron-phonon interaction creates straight one-dimensional tunnels (SODTs) for charge density of electrons and/or holes as revealed by the calculations based on density functional theory (DFT). The SODTs act as resistance-free superhighways and are correlated to the Cooper pairs in the Bardeen-Cooper-Schrieffer (BCS) theory. The formation of SODTs implies that the electron-phonon interaction in the BCS theory can be represented by the difference in *charge densities* between SCC and NCC predicted by DFT. The present work highlights that in conventional superconductors, SODTs are embedded within the bulk materials and are easily destroyed by phonon vibrations, resulting in a low critical superconducting temperature ( $T_C$ ). Conversely, in unconventional superconductors such as $YBa_2Cu_3O_7$ ($YBCO_7$), SODTs are protected by a layered *pontoon* structure with very weak bonding to the bulk materials, maintaining SODTs' stability at higher temperatures and leading to a much higher $T_C$. The present approach is validated for 14 conventional superconductors of 18 pure elements and $MgB_2$ examined in this work, including the presently predicted superconductivity in Cu, Ag, Au, Sb, Bi, and $MgB_2$ at 0 K and 0 GPa, and one unconventional superconductor of $YBCO_7$. Our discovery indicates that DFT can be a practical tool for predicting superconductors, enabling a systematic search for new superconducting materials in the future.

**Keywords:**
Bardeen-Cooper-Schrieffer (BCS) theory; Density functional theory (DFT); Superconducting configuration (SCC); Normal conducting configuration (NCC); Straight one-dimensional tunnels (SODTs); Pure element superconductors; $MgB_2$, $YBa_2Cu_3O_{7-\delta}$

## 1 Introduction

Superconductivity is a phenomenon discovered by Kamerlingh Onnes in 1911 [1] in mercury (Hg) with its electrical resistance vanished at temperatures below a critical temperature ($T_C$) of 4.2 K. The Bardeen-Cooper-Schrieffer (BCS) theory [2], introduced in 1957, provides a microscopic understanding of superconductivity based on the formation of Cooper pairs. These electron pairs, formed via electron-phonon interactions, have lower energy than the Fermi energy and can move freely within the material. However, due to the weak pairing interaction ($\sim 10^{-3}\ eV$), thermal energy can easily disrupt the pairs, leading to conventional superconductors with low $T_C$. In addition to electrons as charge carriers, superconductivity can also occur with holes as charge carriers [3–5].

One significant milestone in superconductivity was the discovery of superconductors with $T_C$ exceeding the limit of 30 K as suggested by the BCS theory. This breakthrough began with $CuLa_{1.85}Ba_{0.15}O_4$ [6] with a $T_C$ of 35 K, and soon advanced to 80 to 93 K for $(Y_{0.6}Ba_{0.4})_2CuO_{4-\delta}$ [7]. Currently, the highest $T_C$ superconductor at ambient pressure in the cuprate family is $HgBa_2Ca_2Cu_3O_{8+\delta}$, achieving $T_C$ values between 133 and 138 K with T1 substitution of Hg [8,9].

Under high pressures, various hydrogen-containing compounds have demonstrated even higher $T_C$ values. For instance, $LaH_{10}$ exhibits superconductivity at 250 K under 170 GPa [10], with ongoing investigations into their Meissner effect [11]. A more recent development in this field is the observation of room-temperature one-dimensional (1D) superconductivity at 300 K in cleaved highly oriented pyrolytic graphite, which features dense arrays of nearly parallel surface line defects [12].

One major theoretical breakthrough in science since the BCS theory is the density functional theory (DFT) [13,14]. DFT provides a solution to the many-body Schrödinger equation in quantum mechanics. It postulates that for any given system, there exists a ground state configuration at 0 K and 0 GPa where the energy is minimized, described by a universal functional of the interacting electron gas density [13]. This unique ground state electron density is determined by separating the independent electron kinetic energy and long-range Coulomb interaction energy, thus transforming the many-body electron problem into one involving independent valence electrons with an exchange-correlation (X-C) functional of the electron density and an associated X-C energy [14]. Currently, DFT plays a central role in predicting $T_C$ of superconductors either through the Eliashberg equation with model parameters or by fully exploring superconductors using DFT (SCDFT) or even beyond, incorporating nonadiabatic effects [15].

The utmost fundamental challenge in DFT for superconductivity is to differentiate the superconducting configurations (SCCs) and the normal conducting configurations (NCCs) at 0 K. The key difference between BCS theory and DFT is that the former requires the explicit consideration of electron-electron interactions in the formation of Cooper pairs, while the latter considers valence electrons independent of each other. Specifically, in DFT, both the electron-electron and electron-phonon interactions are treated indirectly through their contributions to and interactions with the overall potential of the system. Consequently, DFT cannot *directly* simulate the Cooper pairs which require a *direct* description of those interactions. However, it is important to realize that DFT formulated by Hohenberg and Kohn [13] is an exact theory of many-body systems and should be able to differentiate the carrier charge densities of SCCs and NCCs based on the hypothesis presented earlier by one of the present authors [16]. The hypothesis articulates that the electron density due to the formation of Cooper pairs could be predicted by DFT, i.e., the electron density should be identical to each other whether it is created by Cooper pairs or DFT based on formulation by Hohenberg and Kohn [13]. The present work aims to validate this hypothesis by considering the electron-phonon interactions in the BCS theory through perturbation of nuclei positions to induce the electron density (see Section 3). Furthermore it is noted that Lüders et al. [17] developed an approach for the description of superconductors in thermal equilibrium within a formally exact density functional framework, which was applied to the prediction of $T_C$ of pure elements by Marques et al. [18]. More recently, Schmid *et al.* used ab initio low-energy effective Hamiltonians and variational Monte Carlo calculations to study superconductivity order parameters in four carrier doped cuprates [19]. However, the explicit differentiation of SCCs and NCCs at 0 K is not fully addressed.

A key discovery in the recently developed strongly constrained and appropriately normed (SCAN) meta-generalized-gradient approximation (metaGGA) in DFT [20–23] shields light on this challenging topic. In SCAN metaGGA, the strong correlations within a symmetry-unbroken ground-state wavefunction can show up in approximate DFT as symmetry-broken spin densities or total densities due to soft modes of fluctuations such as spin-density or charge-density waves at nonzero wavevector. Consequently, an approximate X-C functional with symmetry breaking, though less accurate than an exact functional, can be more revealing with its utility demonstrated for a number of cases [22–24]. This inspired the present authors to search for SCCs as the symmetry-broken, perturbated configurations of their NCCs.

In the present work, we study the electron-phonon interaction responsible within SCCs by introducing atomic perturbations to NCCs. This results in a *correlated* redistribution of electrons similar to the principles in the SCAN metaGGA. Detailed DFT calculations for SCCs and NCCs are provided in Section 3, followed by results and discussion for 18 pure elements, $MgB_2$, and $YBa_2Cu_3O_{7-\delta}$ ($YBCO_6$ and $YBCO_7$) in Section 4 with 13 pure elements being conventional superconductors including the presently predicted superconductivity in Cu, Ag, Au, Sb, and Bi at 0 K and 0 GPa. Finally, a summary is presented in Section 5.

## 2 Prediction of superconductivity through electron-phonon interactions at 0 K

Based on weak coupling in the BCS theory, the superconducting transition temperature ($T_C$) is commonly evaluated by [25,26],

$$T_C = 0.85\Theta_D e^{-1/n(\varepsilon_F)\phi_{el-ph}} \qquad Eq.\ 1$$

where $\boldsymbol{\Theta_D}$ is the Debye temperature derived from the highest-frequency vibrational mode in the system, $\boldsymbol{n(\varepsilon_F)}$ the electron density of states (eDOS) at the Fermi level, and $\boldsymbol{\phi_{el-ph}}$ an effective electron-phonon attractive interaction [27,28]. This mean field formula is also used in predicting $\boldsymbol{T_C}$ of various hydrides [26,29–32] with strong anharmonicity included [33]. The superconductor must be a conductor with non-zero $\boldsymbol{n(\varepsilon_F)}$ based on Eq. 1.

Currently, the matrix elements of electron-phonon interactions are obtained from the linear response [34] or finite difference methods [35,36]. They have been used to evaluate electron–phonon coupling constant together with DFT energies and phonon frequencies, which is further utilized in the semiempirical McMillan equation to evaluate $T_C$. A parameter called Coulomb pseudopotential was introduced to account for the repulsive electron-electron interaction [37].

Additional considerations have been taken into account for hydrides such as superconducting state and anharmonicity to improve the calculations from the McMillan equation [26].

One key concept in the BCS theory of superconductivity is the superconducting gap that represents the energy gain for two electrons upon formation of a Cooper pair, predicted to be related to $T_C$ as follows for conventional superconductors at 0 K and fell to 0 at $T_C$ [38,39],

$$\Delta(T = 0K) = 1.764k_B T_C \qquad \textit{Eq. 2}$$

The superconducting gap is directly related to the energy difference between SCC and NCC, $\Delta E_{SN}$, commonly referred as condensation energy, as follows [2,40,41]

$$\Delta E_{SN} = -\frac{1}{2}n(\varepsilon_F)\Delta^2 \qquad \textit{Eq. 3}$$

It is worth noting that a negative value of $\Delta E_{SN} < 0$ indicates SCC is the ground state. In contrast, a positive value of $\Delta E_{SN} > 0$ indicates a non-superconducting system.

The present work aims to examine the impact of the electron-phonon interaction on electron redistribution and $\Delta E_{SN}$ and understand its implication on the formation of features that represent superconductivity at 0 K. For pure elements, we investigated the charge density of SCCs through systematic perturbation of atoms in NCCs by DFT-based calculations using the finite difference method [35] to probe the electron-phonon interactions. For $YBCO_6$ and $YBCO_7$, NCCs were created from their SCCs in the present work as detailed in Section 4.4. The fully relaxed SCCs and NCCs were used to determine $\Delta E_{SN}$ though the volume difference is less than 0.03% for pure elements and 0.72% and 0.55% for $YBCO_6$ and $YBCO_7$, respectively. To facilitate the plots of the morphology of the SCC-NCC charge density difference (SNCDD), the equilibrium volume of NCCs is used for both SCC and NCC for pure elements, while that of SCC is used for $YBCO_6$ and $YBCO_7$. As it will be shown in next sections, the straight one-dimensional tunnels (SODTs) are identified as the carrier superhighway to mitigate scattering and correlated with the concept of Cooper pairs at 0 K in the BCS theory.

Our concept for both conventional low temperature and unconventional high temperature superconductors are as follows. In conventional superconductors, SODTs are embedded within the bulk materials and are easily destroyed by phonon vibrations, resulting in low $T_C$. While in unconventional superconductors exemplified by $YBCO_7$, its SODTs form in a layered structure that has very weak bonding with the bulk materials. This layered structure floats in bulk materials much like a *pontoon* floating in water. Consequently, SODTs in $YBCO_7$ can maintain their stability at higher temperature, resulting in much higher $T_C$. Our concept has been validated by DFT-based calculations and available experimental data presented in the next sections.

## 3 Details of DFT-based calculations

To enable DFT-based calculations of superconducting and normal states (SCCs and NCCs) at 0 K, the studied materials are categorized based on their symmetry: (category 1) highly symmetric structures with fixed atomic positions, and (category 2) lower-symmetry structures with variable atomic positions. SCCs and NCCs are constructed accordingly for each category.

For the present materials of study in the first category, we built the 2×2×2 supercells with respect to the crystallographic cells of A1 (i.e., fcc), A4 (i.e., diamond), A6, and A7 lattices of pure elements, as well as the hexagonal compound $MgB_2$, with their unit cells from Materials Project [42], resulting in the 32-, 64-, 36-, 54-, and 24-atom supercells, respectively, as shown in the supplementary Table S 1. They represent their respective NCCs. The atoms in fully relaxed NCCs were perturbed on every other (001) layer as follows,

$$[x_0 + n_x \Delta d_{ini} \quad y_0 + n_y \Delta d_{ini} \quad z_0 + n_z \Delta d_{ini}] \qquad \textit{Eq. 4}$$

where $x_0$, $y_0$, $z_0$ are the coordinates of atoms in NCC in cartesian coordinate system; $n_x$, $n_y$, and $n_z$ are random number 0, 1, or -1; and $\Delta d_{ini}$ is the initial perturbation amplitude from 0.1 Å to 0.7 Å to ensure that the perturbated atoms do not return to their original positions. It is important to emphasize that, while these amplitudes exceed those commonly used in harmonic phonon and electron-phonon coupling calculations [43,44], they are essential for identifying SCCs as well as mimicking the effects of strong electron-phonon interactions. After testing calculations using different initial $\Delta d_{ini}$ values along x-, y-, and/or z-axis direction, the configuration yielding the lowest total energy is selected as the SCC. The adopted $\Delta d_{ini}$ values are listed in supplementary Table S 2 along with the representative structure files listed in supplementary Table S 1.

For the present materials of study in the second category, we employed a 2×2×1 supercell with 48 atoms for $YBCO_6$ and 52 atoms for $YBCO_7$, where the undistorted, symmetry-unbroken YBCO configurations were built by adjusting the atoms on the Cu2-O2-O3 plane to the same z level. More details are given in Section 4.4 with their structure files listed in supplementary Table S 1 and the plots in Figure 4 and Figure S 22.

All the present DFT-based calculations were performed by VASP code [45]. The ion-electron interaction was described by the projector augmented wave (PAW) method [46]. Two X-C functionals were employed, i.e., the widely used functional of GGA-PBE [47] and one of most accurate metaGGA functionals of $r^2$SCAN [20,48]. It is important to note that none of the existing X-C functionals can perfectly capture the energy difference $\Delta E_{SN}$, as discussed in Section 4 and the Summary. In VASP calculations, electron configurations for each element were the same as those used by the Materials Project [42] with their valance electrons and other settings shown in Table S 2. The energy convergence criterion of the electronic self-consistency was at least $10^{-6}$ eV/atom for all calculations. Convergence tests regarding *k*-point meshes and plane wave cutoff energy ($E_{cut}$) were performed, and two of them are shown in Figure S 1 for pure element Al using $r^2$SCAN. It indicates that the predicted energy difference, $\Delta E_{SN}$, between SCC and NCC, i.e., the condensation energy in the BCS theory shown by Eq. 3, becomes convergent when the *k*-point meshes are larger than (6×6×6), and the $E_{cut}$ value has less impact. In the present work, the selected *k*-point meshes were (7×7×7) or (9×9×9), and the $E_{cut}$ values were determined by VASP using the setting of PREC = High for pure elements and $MgB_2$ with their values shown in Table S 2.

For $YBCO_6$ and $YBCO_7$, the automatic *k*-point meshes were generated to sample the Brillouin zone in terms of the assigned $R_k$ value of 35 to determine the subdivisions of *k*-point meshes, and $E_{cut}$ = 520 eV was used for final calculations as shown in Table S 2. Phonon calculations of fcc Al and $YBCO_7$ were performed by the supercell approach [36] and GGA-PBE in terms of the YPHON code [49], with VASP to calculate force constants [50] by means of the finite difference method with $E_{cut}$ = 400 eV and $R_k$ = 25 for $YBCO_7$ and $E_{cut}$ = 400 eV and *k*-meshes = (7×7×7) for Al. Note that the GGA-PBE predicts nonmagnetic (NM) configurations for YBCO [42], while the $r^2$SCAN predicts the ground states of G-type antiferromagnetic (AFM) configurations as pointed out by Zhang et al. [51]. In the present work, the NM configurations of YBCO were used for PBE calculations while the AFM configurations were used for $r^2$SCAN calculations.

In the present calculations, the Methfessel-Paxton technique [52] was used for structural relaxations and phonon calculations, and the tetrahedron method with a Blöchl correction [53] was used to calculate charge density. The minimum and the maximum charge density differences and the isosurface levels ($F_{level}$) to plot SNCDD using the VESTA code [54] are listed in Table S 2. Note that the $F_{level}$ is chosen as the approximate maximum value that still allows the SNCDD to appear just connected in the visualization. Selecting lower $F_{level}$ values would result in a fully connected SNCDD, forming continues 3D tunnels across all materials. Equilibrium properties of YBCO at 0 K and 0 GPa, including the equilibrium volume ($V_0$), bulk modulus ($B_0$) and its derivative with respect to pressure ($B'$), were fitted by the four-parameter Birch-Murnaghan equation of state (EOS) [55] with inputs from DFT-based energy versus volume data points.

Due to the proposed symmetry breaking in SCCs, symmetry constraints were fully disabled in the present VASP calculations for all pure elements and $MgB_2$, except for YBCO (including both $YBCO_6$ and $YBCO_7$ with their symmetries reported in supplementary Table S 3).

## 4 Results and discussion

Pure elements are used to search for SCCs due to their simplicity and available experimental data in the literature [56]. Table 1 summarizes the present results of 18 pure elements, $MgB_2$, and YBCO from DFT-based calculations using GGA-PBE [47] and metaGGA-$r^2$SCAN [21], including $\Delta E_{SN}$, SNCDD, and predicted superconductivity in comparison with available experiments [56–58]. As outlined in Section 2, the present criteria for classifying a material as a superconductor at 0 K include: (i) its metallic or conducting nature, (ii) the formation of SODTs, and (iii) the SCC representing the ground state. However, the third criterion is currently omitted due to the limitations of existing X-C functionals in accurately resolving the energy difference between SCC and NCC states. For pure elements, SNCDDs due to electrons and holes are similar so only electron SNCDD for pure elements are presented in the main text, and hole SNCDDs are included in supplementary material. For YBCO, both electron and hole SNCDDs are presented in the main text or the supplementary material.

## 4.1 Pure metal elements with fcc (A1) structure

Figure 1 and Figure S 2 show the SNCDDs of fcc Al with Figure 1and Figure S 2a by PBE, and Figure S 2b by $r^2$SCAN, respectively. Both electron and hole SNCDDs show the formation of SODTs along [110] direction. It can be seen that $r^2$SCAN predicts that the SCC of Al is a ground state with the predicted energy difference $\Delta E_{SN}$ = -1.114 meV/atom, while PBE shows that NCC is more stable with $\Delta E_{SN}$ = 0.076 meV/atom as shown in Table 1. This difference can be attributed to the approximations in current X-C functionals, while the values in the literature [41] are in the range of $10^{-6}$ meV/atom, much smaller than the DFT accuracy. Nevertheless, the existence of SODTs is verified for Al and other superconducting elements as shown below.

Experimentally, $T_C$ of bulk Al is about 1.18 K at 0 GPa and reducing to 0.075 K at 6.2 GPa [59]. In addition, Singh et al. [60] reported the measured $T_C$ = 1.7 K (or 1.9 K) using a 80- (or 30-) nm single crystal Al nanowire with its [110] as the preferred growth direction. This $T_C$ is higher than the 1.18 K for bulk Al [59], implying that [110] of Al is a preferred direction of superconductivity in accordance with the direction of SODTs predicted in the present work. The present results of Al are summarized in Table 1 along with experimental information both showing fcc Al as a superconductor at 0 K and 0 GPa.

It is observed that atomic bonding behaviors in both NCC and SCC of fcc Al are quite similar, accounting for its low $T_C$ temperature. For example, *Figure S 3* shows the predicted stretching force constants (SFC's) from phonon calculations for fcc Al in terms of the 32-atom NCC and SCC, respectively, at an external pressure of 0 GPa. It can be seen that the fluctuation of bond lengths in SCC has a very small standard derivation $\delta$ = 0.00053 Å for the first nearest neighbors around 2.856 Å. Correspondingly, the fluctuation of SFC's in SCC has a very small standard derivation $\delta$ = 0.0026 eV/Å$^2$ around 1.31 eV/Å$^2$. *Figure S 4* depicts phonon dispersions in NCC and SCC plotted using the 1-atom primitive cells or the 32-atom supercells, respectively. It shows that the dispersion curves of SCC are disturbed with respect to those of NCC due to symmetry breaking, making some degenerate curves separated, such as the acoustic branches from Γ to R. Unlike the differences observed in phonon dispersions, the difference in electronic structures in NCC and SCC is negligible as shown in the predicted band structures and electron density of states for fcc Al in *Figure S 5*.

Figure 2 illustrates the PBE predicted SNCDD of fcc Pb, showing SODTs along [110] direction; see also the hole SNCDDs by PBE and electron SNCDDs by $r^2$SCAN in Figure S 6. Table 1 shows that the $\Delta E_{SN}$ values are close to zero ($|\Delta E_{SN}|$ < 0.013 meV/atom) in terms of both PBE and $r^2$SCAN. Based on the predicted SODTs, we conclude that Pb is a superconductor at 0 K and 0 GPa. Experimentally, bulk Pb has a measured $T_C$ = 7.2 K [56], and He et al. [61] showed that Pb nanowire has an enhanced $T_C$ which is 3-4 K above the bulk $T_C$. The textures of Pb nanowire include $\langle 200 \rangle$, $\langle 110 \rangle$, and $\langle 123 \rangle$ [62], implying that $\langle 110 \rangle$ is among the preferred superconducting direction as predicted by the direction of SODTs in the present work. These experimental observations along with the present DFT predictions are summarized in Table 1.

The present results by PBE and $r^2$SCAN indicate that most fcc elements have the similar SNCDD features as those of Pb, including the group IB elements of Cu, Ag, and Au and the group VIII elements of Rh, Ir, Pd, and Pt as shown in Figure S 7 to Figure S 13. These 7 fcc elements form SOSTs and are all superconductors at 0 K and 0 GPa based on our theory. The superconductivity in Rh, Ir, Pd, and Pt has been reported in the literature at ambient pressure, i.e., $T_C$ = 35 μK for Rh, $T_C$ = 0.1 K for Ir, $T_C$ = 3.2 K for Pd [56], and $T_C \approx$ 1 mK (0.62 ~ 1.38 mK) for Pt [58]. On the other hand, the superconductivity in Cu, Ag, and Au has only been estimated by extrapolation from $T_C$ of fcc alloys rich in noble metals by Hoyt and Mota [63] as $7\times10^{-10}$ K, $8\times10^{-10}$ K, and $2\times10^{-4}$ K (0.2 mK), respectively. However, Hoyt et al. [64] did not observe superconductivity in a polycrystalline sample of 99.9999% Au at 0.22 mK, probably due to its slight higher value than 0.2 mK or lower $T_C$ than the extrapolated value. Notably Ummarino and Zaccone [65] predicted that ultrathin films of Au, Ag, and Cu with suitable thickness (e.g., 0.5 nm) could be superconductors at low but experimentally accessible temperatures.

For alkali earth elements, SNCDD plots in Figure S 14 for Ca and Figure S 15 for Sr show that PBE predicts SODTs, however, $r^2$SCAN predicts 3D networks for both Ca and Sr. We hence suggest that Ca and Sr are not superconductors at 0 K and 0 GPa. Experimentally the superconductivities were only observed at high pressures for Ca and Sr based on the review work by Buzea and Robbie [56] and Hamlin [66] as shown in Table 1, and neither is superconducting at 0 GPa.

### 4.2 Pure elements with A4 structure (Si, Ge, and Sn)

Both Si and Ge in the A4 diamond structure are semiconductors at ambient pressure but superconductors at high pressures with different structures [67], such as $T_C$ = 8.5 K at 12 GPa for Si and $T_C$ = 5.4 K at 11.5 GPa for Ge, both in the β-tin structure [56]. The present DFT calculations with both PBE and $r^2$SCAN predict Si as a semiconductor and Ge as a conductor in agreement with other DFT predictions [42]. Figure S 16 shows that SNCCD of Si forms SODTs by both PBE and $r^2$SCAN. Figure S 17 shows that SNCCD of Ge forms SODTs by $r^2$SCAN but 3D networks by PBE. Our theory thus indicates that the semiconductors Si and Ge are not superconductors at 0 K and 0 GPa, due to the lack of free electrons at their Fermi levels. However, it is less certain for Ge due to the formation of SODTs as predicted by $r^2$SCAN.

α-Sn is a post-transition metal in the A4 structure and exhibits superconductivity at ambient pressure with $T_C$ = 3.7 K [56]. Figure S 18 shows that its SNCCD forms SODTs, and the predicted $\Delta E_{SN}$ values between SCC and NCC are -0.270 and -0.102 meV/atom by both PBE and $r^2$SCAN, respectively, as shown in Table S 2, indicating that α-Sn is a superconductor at 0 K and 0 GPa, in agreement with experimental observation [56] as shown in Table 1.

### 4.3 Pure elements with A6 and A7 structures (In, As, Sb, and Bi)

In is a post-transition metal in the A6 structure. Figure S 19 depicts that its SNCCD forms SODTs by both PBE and $r^2$SCAN. Table S 2 shows that the SCCs are ground state with $\Delta E_{SN}$ = 0.061 meV/atom by PBE and -0.328 meV by $r^2$SCAN, indicating that In is a superconductor at 0 K and 0 GPa. Experimentally, the measured $T_C$ was 3.4 K for bulk In at ambient pressure [56], agreeing with the present DFT results as shown in Table 1.

Arsenic (As) is a metalloid in the A7 structure. DFT predicts the pronounced zigzag 1D tunnels as shown in Figure 3, which scatter migrating electrons and holes. The formation of such zigzag 1D tunnels is a direct consequence of lattice symmetry, as As adopts a distorted rhombohedral structure driven by Peierls distortion [68], implying that this heavy zigzag-type lattice is generally unfavorable for superconductivity, as it impedes coherent charge transport. The predicted $\Delta E_{\mathrm{SN}}$ values are about -0.096 meV/atom by PBE (or -14.661 meV/atom by r$^2$SCAN due to large volume difference between SCC and NCC; see details in Table 1). As is hence not a superconductor at 0 K and 0 GPa based on our theory. Experimentally, bulk As was observed with $T_C$ = 0.1 to 2.7 K at 13-24 GPa [56,67].

Both Sb and Bi have the A7 structure at low temperatures. Their SNCDDs are plotted in Figure S 20 and Figure S 21, respectively, showing the formation of SODTs by both PBE and r$^2$SCAN. Both Sb and Bi are conductors and with small $|\Delta E_{\mathrm{SN}}|$ values (< 0.054 meV/atom) as shown in Table 1 and Table S 2. Based on our theory, they are both superconductors at 0 K and 0 GPa. Experimentally, Sb and Bi are both superconductors at high pressures, i.e., $T_C$ = 3.6 K at 8.5 GPa for Sb and $T_C$ = 8.7 K at 9 GPa for Bi, respectively [56]. The observed $T_C$ in Bi is as follows: 6.5 – 7.0 K in 3.7 – 4.3 GPa and 6.7 K at 6.8 GPa and then decreases with pressure to 6.0 K at 20-25 GPa with its structures being Bi-III (tetragonal) and Bi-IV (body-centered tetragonal), potentially other structures at higher pressure [67]. Sb transitions to a monoclinic structure around 8 GPa at room temperature and maintains similar $T_C$ = 3.4 K at 15 GPa [67]. Their superconductivity at ambient pressure has not been reported, and our theoretical predictions probably reflect the local structure resembling the metastable configuration thus with very low $T_C$; see also the discussion for Au, Ag, and Cu in Section 4.1.

### 4.4 Conventional superconductor $MgB_2$

Magnesium diboride ($MgB_2$) is an inorganic compound whose superconducting properties were discovered in 2001 [69]. With a critical temperature $T_C = 39\ K$ under ambient conditions, it holds the highest $T_C$ amongst conventional superconductors. Figure S 29 presents the predicted electron and hole SNCDDs obtained using the r$^2$SCAN functional, revealing the formation of SODTs along c-axis for electrons and b-axis for holes. Furthermore, the present DFT calculations using both PBE (GGA) and r$^2$SCAN confirm that $MgB_2$ is a conductor with nonzero electron density at the Fermi level. The computed energy difference $\Delta E_{\mathrm{SN}}$ is negative, specifically, -0.030 meV/atom by PBE and -0.335 by r$^2$SCAN, indicating the SCC is energetically favorable. Based on the criteria established in the present work, $MgB_2$ is hence classified as a superconductor, as summarized in Table 1.

### 4.5 High-temperature superconductor of $YBa_2Cu_3O_{7-\delta}$

$YBCO_6$ is an insulator and becomes a conductor at $YBCO_{6.5}$ and a superconductor at $YBCO_{6.93}$ with $T_C \approx 93\ K$ and at $YBCO_7$ with $T_C \approx 88\ K$ [70]. Table S 3 shows the presently predicted lattice parameters and atomic positions of $YBCO_6$ and $YBCO_7$ by PBE, which are in good agreement with experiments [71,72]. For example, the measured and the predicted (in parentheses) lattice parameters for $YBCO_7$ are $a$ = 3.820 (3.837) Å, $b$ = 3.886 (3.919) Å, and $c$ = 11.684 (11.869) Å, at room temperature (0 K), respectively. Table S 4 lists the predicted equilibrium properties ($V_0$, $B_0$, and $B'$) of $YBCO_6$ and $YBCO_7$ by EOS fittings in terms of both PBE and $r^2$SCAN, which are also in good agreement with available measurements [71]. For example, the predicted $B_0$ at 0 K (115.5 GPa by $r^2$SCAN) agrees well with the measured 115 GPa of $YBCO_7$ at room temperature by high-pressure X-ray diffraction [73].

Figure 4(a) shows the fully relaxed 2×2×1 supercell of $YBCO_7$, i.e., the SCC, by PBE, illustrating that the Cu1-O1 plane is flat, while the Cu2-O2-Cu2-O3 plane is rather wavy in accordance with computational predictions and experimental observations in the literature [71,72]. The atomic positions are shown in Table S 3. The stretching force constants (SFCs, see detailed methodology in [50]) obtained from phonon calculations by PBE are plotted in Figure 4 (b). The SFC between Cu1-O4 is the largest, followed by those of Cu2-O2, Cu2-O3, and Cu1-O1, while the SFC of Cu2-O4 is negative (-1.6 eV/Å$^2$), and so are the SFCs of Ba-O2 and Ba-O3 (-0.5 eV/Å$^2$) with a long bond length about 3 Å. Bonding strengths represented by these SFCs [50] indicate two frames in $YBCO_7$ with the frame 1 being the Cu1-O4-Ba-O1 structure at the top and bottom of Figure 4 (a) and the frame 2 being the Cu2-O2-O3-Y structure in the middle of Figure 4 (a). The SFCs within both frames are large (> 3 eV/Å$^2$), while the SFCs between them are small (< 1 eV/Å$^2$). It is further noted that the SFCs of Y-O2 and Y-O3 (< 0.9 eV/Å$^2$) are much smaller than those of Cu2-O2 and Cu2-O3 (> 4 eV/Å$^2$) in the frame 2, thus loosely bonding the top and bottom layers within the frame 2 and with minimal disturbance on the Cu2-O2 and Cu2-O3 bonding which are responsible for the formation and stability of SODTs in the superconductor.

The crystallographic information presented in Table S 3 depicts that the rigid frame 1 is symmetry-unbroken with Cu1-O1 on the same x and z levels (along a- and c-axis, respectively) and Cu1-O4 on the same x and y levels (along a- and b-axis, respectively), while the wavy frame 2 is symmetry-broken with O2-O3 on the same z level and Cu2 shifting towards the frame 1 and Y loosely connecting the two O2-Cu2-O3 layers. The frame 2 structure in the middle of Figure 4 (a) thus resembles a three-layer pontoon structure floating between the two rigid frame 1 structures. Using these characteristics, we build the undistorted, symmetry-unbroken $YBCO_7$, i.e., its NCC, by constraining the atoms in the Cu2-O2-Cu2-O3 plane on the same z level as shown in Figure S 22(a). Its structure file is provided in supplementary Table S 1.

Similarly to $YBCO_7$, we built the undistorted $YBCO_6$ by constraining the atoms in the Cu2-O2 plane on the same z level with its structure file provided and listed in Table S 1. Figure S 23 (by PBE) shows that the double 2D tunnels connected by Y atoms and parallel to a-b plane are formed by O2 atoms for electron SNCDD and by Cu2-O2 atoms for hole SNCDD, but without SODTs. Figure S 24 (by $r^2$SCAN) shows similar 2D tunnels for both the electron and hole SNCDDs without SODTs. $YBCO_6$ is thus not a superconductor due to the lack of SODTs based on our theory, in agreement with experimental observations [70].

Figure 5(a) and Figure S 25(a) by PBE show the electron SNCDD in $YBCO_7$ with 2D tunnels parallel to the a-b plane between the Ba-O4 and the Cu2-O2-Cu2-O3 planes, while Figure 5(b) and Figure S 25(b) depict the hole SODTs parallel to a-axis along the Cu2-O2 atoms. The $r^2$SCAN predicts similar but slightly different behaviors (c.f., Figure S 26), indicating that the electron SNCDD in $YBCO_7$ forms SODTs parallel to b-axis and between the Ba-O4 and the Cu2-O2-Cu2-O3 planes, while the hole SNCDD forms 2D tunnels parallel to the a-b plane. Our theory thus indicates the superconductivity in $YBCO_7$, suggesting that the superconductivity is dominated by electron conduction (by $r^2$SCAN) or hole conduction (by PBE) in the b-axis or a-axis direction, respectively, in agreement with the Hall measurements by Bauhofer et al. [74]. Bauhofer et al. [74] also reported the anisotropic $T_C$ values in single crystal $YBCO_{6.9}$ using the measured critical fields and the higher $T_C$ in its a-b plane, supported by the current predictions with SODTs along the a-axis or b-axis in the a-b plane.

While the electronic structures and eDOS of SCCs and NCCs in low $T_C$ superconductors such as Al are very similar to each other as shown in *Figure S 5*, eDOS of SCC and NCC for $YBCO_7$ are clearly different as depicted in Figure S 27 due to the large fluctuations of bond lengths (Table S 3) and energy differences (-30 ~ -40 meV/atom, c.f., Table 1). Particularly, one of the two peaks just above the Fermi level in eDOS of NCC, depicted in Figure S 27(a), changes to two smaller peaks in eDOS of SCC as shown in Figure S 27(b), which is enlarged in Figure S 28(a). The integrated eDOS is plotted in Figure S 28(b), showing more electrons in NCC than that in SCC in the range of 0.2 ~ 0.6 eV above the Fermi level, indicating the formation of SODTs in SCC of $YBCO_7$ lowers the Fermi energy likely related to the formation of energy gap as indicated by the BCS theory [75].

## 5 Summary

The BCS theory posits that superconductivity occurs due to the formation of Cooper pairs through electron-phonon interactions. The present work systematically investigates the electron-phonon interactions at 0 K in 18 pure elements, $MgB_2$, $YBCO_6$, and $YBCO_7$ by DFT-based calculations through atomic perturbations in their conventional ground state configurations. By plotting the charge density difference between the perturbed and unperturbed configurations, it is discovered that the formation of SODTs in a conductor correlates with its superconductivity for both conventional and unconventional superconductors. It is concluded that these SODTs enable scattering-free migration of electrons and/or holes, i.e., resistance-free superhighways for migration of electrons or holes. The SODTs are likely related to Cooper pairs in the BCS theory. Among the 18 pure elements and $MgB_2$, both electron and hole SODTs are observed in Al, Pb, Cu, Ag, Au, Rh, Ir, Pd, and Pt in the fcc structure, Sn in the A4 structure, In in the A6 structure, Sb and Bi in the A7 structure, and $MgB_2$ in the hexagonal structure with their superconductivity observed experimentally at 0 K and 0 GPa except for Cu, Ag, Au, Sb, and Bi probably due to their extremely low $T_C$ values. While Ca and Sr in the fcc structure, Si and Ge in the A4 structure, and As in the A7 structure do not possess superconductivity at 0 K and 0 GPa due to the lack of SODTs or nonconductors in agreement with experimental observations. In YBCO, both electron and hole SODTs are observed for unconventional superconductor $YBCO_7$, but not in non-superconductor $YBCO_6$, also in agreement with experimental observations.

It is observed that current DFT calculations struggle to reliably distinguish the energy difference between SCCs and NCCs due to approximations in current X-C functionals, demanding further improvement. Based on the symmetry-broken configurations observed in the development of SCAN, one potential approach is to employ the exact functional for the ground state configuration followed by the zentropy theory [16,76–78] to account for other symmetry-broken configurations. Moreover, once superconducting configurations (SCCs) are identified for a given material, it becomes feasible to predict the superconducting temperature $T_C$ by analyzing the competition between SCC and NCC states. This can be achieved using theoretical frameworks such as the zentropy theory, which represents an ongoing direction in our research on superconductivity.

## 6 Acknowledgements

This work was supported by the U.S. Department of Energy (DOE) through Grant No. DE-SC0023185 and the Endowed Dorothy Pate Enright Professorship at College of Earth and Mineral Science at the Pennsylvania State University (PSU). First-principles calculations were performed partially on the Roar supercomputers at PSU's Institute for Computational and Data Sciences (ICDS), partially on the resources of the National Energy Research Scientific Computing Center (NERSC) supported by the DOE Office of Science User Facility operated under Contract no. DE-AC02-05CH11231 under NERSC Award No. ALCC-ERCAP0022624, and partially on the resources of the Extreme Science and Engineering Discovery Environment (XSEDE) supported by NSF with Grant no. ACI-1548562. We thank Dale Gaines for pointing out the convergent issues of total energy with respect to *k*-point meshes. ZKL thank John Perdew for explaining the symmetry breaking in SCAN during ZKL's visit to Temple University in March 2023 and his interest in the present work through many email communications since. ZKL is also grateful to Gareth Conduit for valuable discussions on superconductivity during the sabbatical leave at University of Cambridge in 2022 and Darrell Schlom for many stimulating discussions.

## 7 Data availability statement

All data and plots that support the findings of the present work are included within this article and its supplementary files.

## 8 ORCID IDs

Zi-Kui Liu https://orcid.org/0000-0003-3346-3696

Shun-Li Shang https://orcid.org/0000-0002-6524-8897

## 9 Table 1

Table 1. DFT-based results to determine superconductivity of pure elements, $MgB_2$, and YBCO (Yes by Y for superconductivity and No by N for no superconductivity), including conductivity (Cond: conductor by Y and nonconductor by N), SCC as ground state (GS: Y for yes and N for no, and the predicted energy difference between SCC and NCC: $\Delta E_{SN}$ in meV/atom), the morphology of the SCC-NCC charge density difference (SNCDD), and the calculated and experimental superconductivity (Yes by Y and not by N), where the settings to plot electron SNCDD are given in Table S 2.

| Mater. | X-C | DFT results | | | Characteristics of SNCDD by DFT | Superconductivity | | Figures |
|---|---|---|---|---|---|---|---|---|
| | | Cond | GS | $\Delta E_{SN}$ | | Calc. [a] | Expt. [b] | |
| Al | PBE | Y | N | 0.076 | SODT along [110] | Y | $Y_0$ | Figure 1 |
| | $r^2$SCAN | Y | Y | -1.114 | | | 1.18 K | Figure S 2 |
| Pb | PBE | Y | Y | -0.013 | SODT along [110] by PBE and [101] by $r^2$SCAN | Y | $Y_0$ | Figure 2 |
| | $r^2$SCAN | Y | N | 0.009 | | | 7.2 K | Figure S 6 |
| Cu | PBE | Y | N | 0.071 | SODT along [110] | Y | | Figure S 7 |
| | $r^2$SCAN | Y | Y | -0.347 | | | | |
| Ag | PBE | Y | Y | -0.025 | SODT along [101] | Y | | Figure S 8 |
| | $r^2$SCAN | Y | Y | -0.109 | | | | |
| Au | PBE | Y | Y | -0.002 | SODT along [101] | Y | | Figure S 9 |
| | $r^2$SCAN | Y | Y | -0.003 | | | | |
| Rh | PBE | Y | N | 0.019 | SODT along [011] | Y | $Y_0$ | Figure S 10 |
| | $r^2$SCAN | Y | Y | -0.001 | | | 35 μK | |
| Ir | PBE | Y | Y | -0.038 | SODT along [110] | Y | $Y_0$ | Figure S 11 |
| | $r^2$SCAN | Y | N | 0.003 | | | 0.1 K | |
| Pd | PBE | Y | N | 0.005 | SODT along [110] by PBE and [011] by $r^2$SCAN | Y | $Y_0$ | Figure S 12 |
| | $r^2$SCAN | Y | N | 0.008 | | | 3.2 K | |
| Pt | PBE | Y | N | 0.002 | SODT along [101] | Y | $Y_0$ | Figure S 13 |
| | $r^2$SCAN | Y | Y | -0.034 | | | ~1 mK | |
| Ca | PBE | Y | Y | -0.0002 | SODT along [011] | N | $Y_h$ | Figure S 14 |
| | $r^2$SCAN | Y | Y | -0.563 | Tunnels formed in 2D on (111) planes | | 15 K @ 150 GPa | |
| Sr | PBE | Y | Y | -0.012 | SODT along [011] | N | $Y_h$ | Figure S 15 |
| | $r^2$SCAN | Y | Y | -0.001 | Tunnels in 2D and 3D | | 4 K @ 50 GPa | |
| Si | PBE | N | Y | 0.000 | SODT along [011] by PBE and [101] by $r^2$SCAN | N | $Y_h$ | Figure S 16 |
| | $r^2$SCAN | N | Y | 0.001 | | | 8.5 K @ 12 GPa | |
| Ge | PBE | Y | N | 0.001 | 3D tunnel | N | $Y_h$ | Figure S 17 |
| | $r^2$SCAN | N | N | 1.073 | SODT approximately along $[\bar{1}01]$ | | 5.4 K @ 11.5 GPa | |
| Sn | PBE | Y | Y | -0.270 | SODT approximately along $[\bar{1}01]$ | Y | $Y_0$ 3.2 K | Figure S 18 |
| | $r^2$SCAN | Y | Y | -0.102 | SODT approximately along $[1\bar{1}0]$ | | | |
| In | PBE | Y | Y | 0.061 | SODT approximately along $[\bar{1}\bar{1}1]$ | Y | $Y_0$ 3.4 K | Figure S 19 |
| | $r^2$SCAN | Y | Y | -0.328 | SODT approximately along $[\bar{1}11]$ | | | |
| As | PBE | Y | N | -0.096 | Pronounced zigzag 1D-type tunnel, along [111] and [110]. Not facile for carrier transfer. | N | $Y_h$ 2.7 K @ 24 GPa | Figure 3 |
| | $r^2$SCAN | Y | N | -14.661[c] | | | | |
| Sb | PBE | Y | Y | -0.018 | SODT along [011] | Y | $Y_h$ 3.6 K @ 8.5 GPa | Figure S 20 |
| | $r^2$SCAN | Y | N | -0.035 | SODT approximately along [001] | | | |
| Bi | PBE | Y | Y | -0.054 | SODT approximately along [010] | Y | $Y_h$ 8.7 K @ 9 GPa | Figure S 21 |
| | $r^2$SCAN | Y | N | 0.024 | SODT approximately along [010] | | | |
| $YBCO_6$ | PBE | Y | N | -7.54 | Double 2D tunnels | N | N | Figure S 23 |
| | $r^2$SCAN | Y | Y | -15.41 | Double 2D tunnels | | | Figure S 24 |
| $YBCO_7$ | PBE | Y | N | -37.79 | Hole SODT along *a*-axis | Y | $Y_0$ ~ 88 K | Figure 5<br>Figure S 25 |
| | $r^2$SCAN | Y | Y | -29.96 | Electron SODT along *b*-axis | | | Figure S 26 |
| $MgB_2$ | PBE | Y | Y | -0.030 | SODT along *c*- or *b*-axis. | Y | $Y_0$ 39 K | Figure S 29 |
| | $r^2$SCAN | Y | Y | -0.335 | | | | |

[a] This work with “Y” (yes) for exhibiting superconducting behavior and “N” (no) for not, based on the present criteria. It should be noted that the critical superconducting temperature $T_C$ is not computed in the present work.

[b] Experimentally observed superconducting elements at ambient pressure (marked by $Y_0$) or high pressure (marked by $Y_h$), and the values indicate the measured $T_c$ [56,57]. The values of Pt, $0.62 \le T_c \le 1.38$ mK, were measured using the compacted, high purity Pt powder [58]. Experimental data of $YBCO_6$ and $YBCO_7$ were taken from [70], and $MgB_2$ from [69].

[c] This large $\Delta E_{SN}$ value of As by $r^2$SCAN is due to large volume change between SCC and NCC after relaxations (up to 25%), a further investigation is needed using such as the energy-volume EOS fitting.

## 10 Five figures for Al, Pb, As and $YBCO_7$

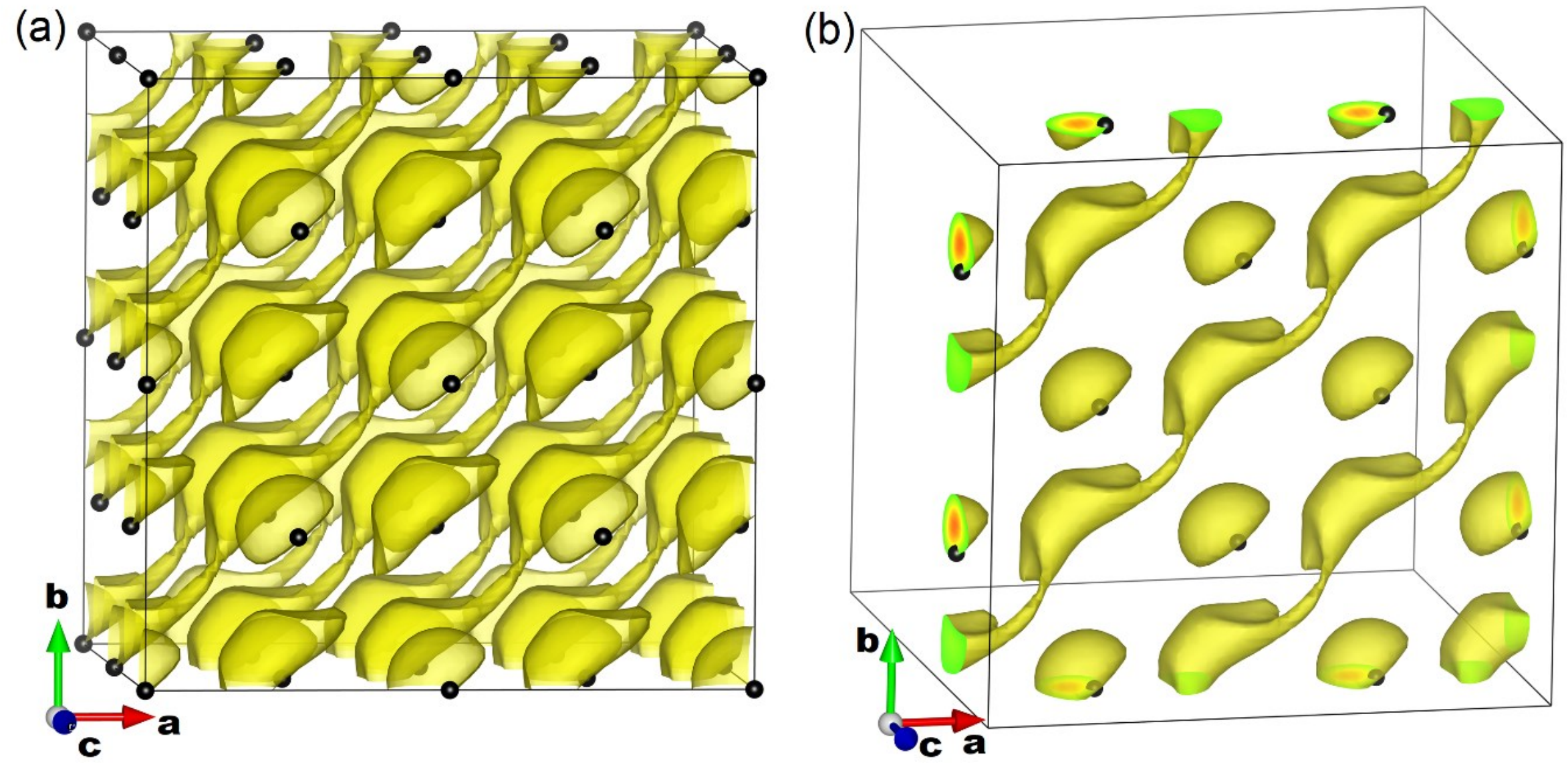

*Figure 1. PBE predicted electron SNCDD in the 32-atom supercell of Al (a, in yellow) and showing partial charge results (b), showing SODTs formed along [110].*

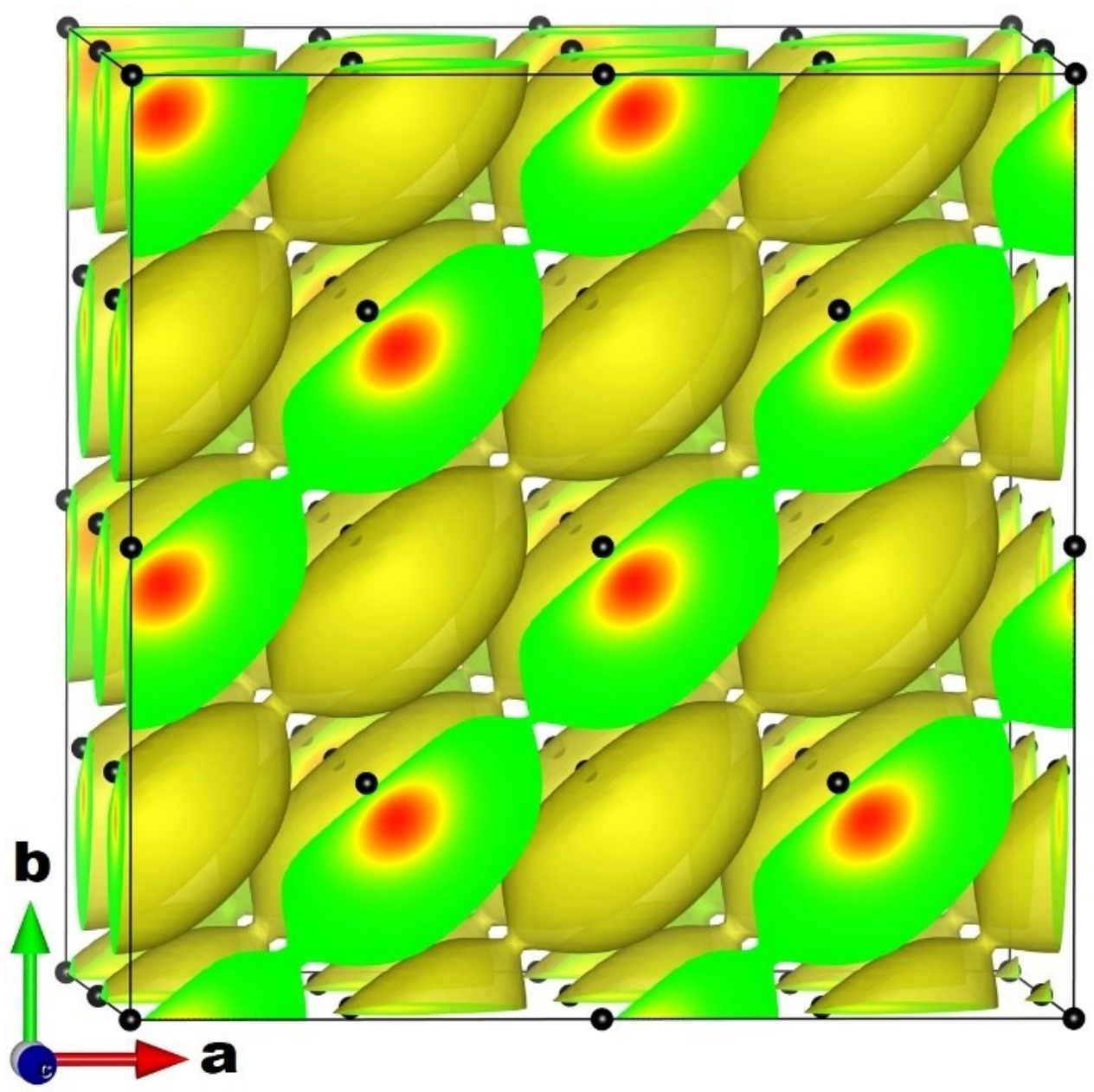

*Figure 2. PBE predicted electron SNCDD in the 32-atom supercell of Pb (in yellow, with showing cut sections) with SODTs formed along [110].*

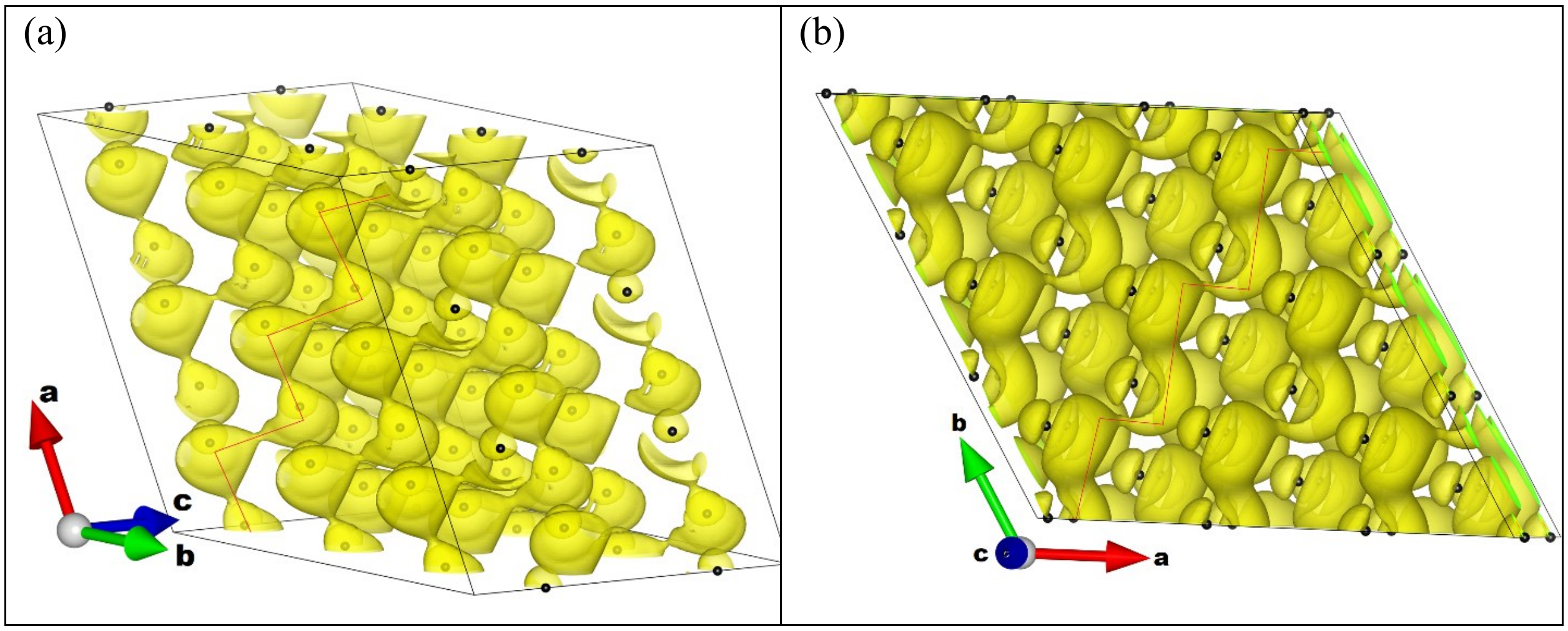

*Figure 3. PBE predicted electron SNCDD (a, in yellow) and $r^2$SCAN predicted electron SNCDD (b, in yellow) of arsenic (As). The red lines indicate one of the pronounced zigzag 1D tunnels formed close to [111] and [110] directions, which are not facile for carrier transfer.*

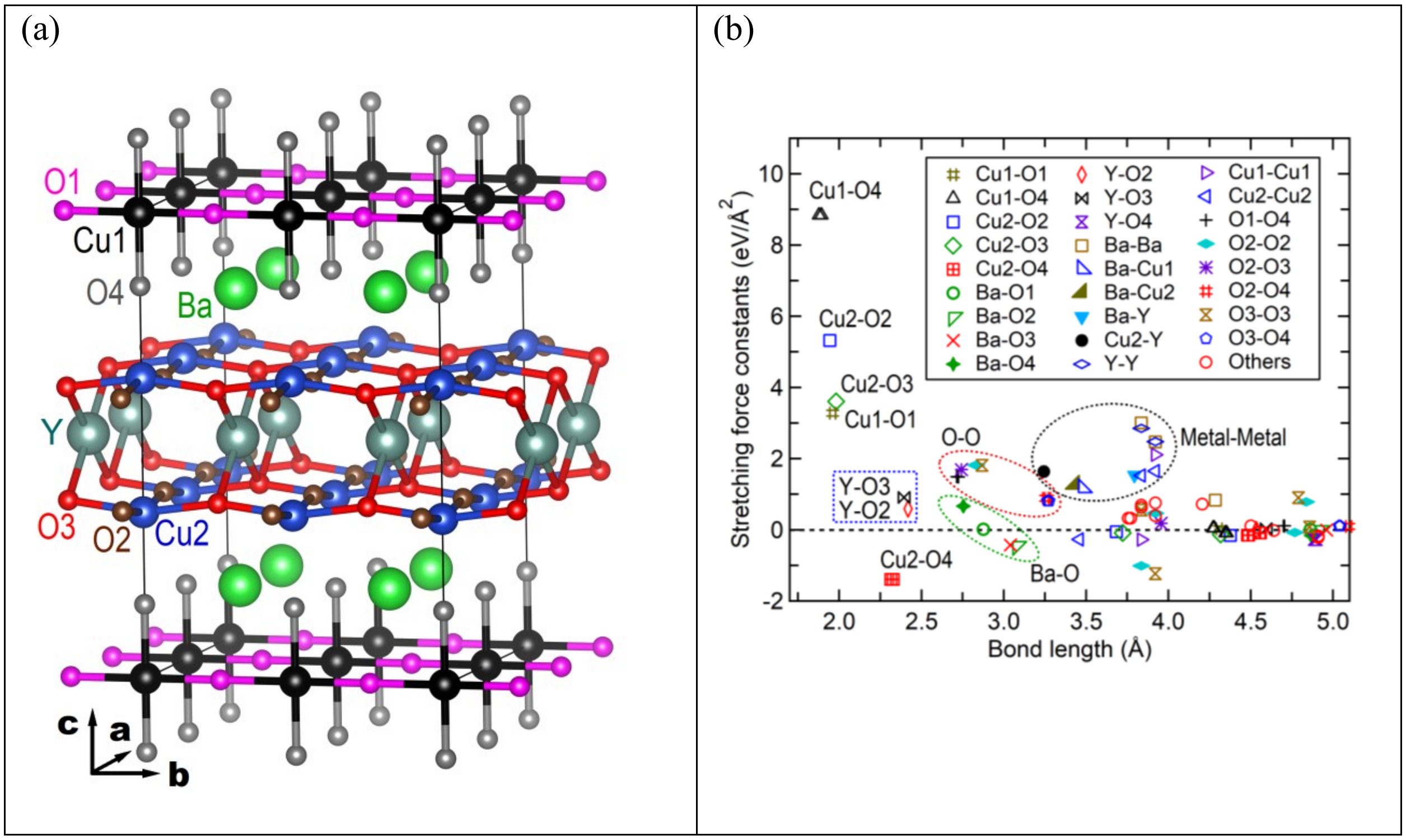

*Figure 4. (a) SCC configuration of the 2×2×1 $YBCO_7$ supercell with the bonds connecting key interactions indicated by (b) the stretching force constants (SFCs) from phonon calculations by PBE. Crystallographic details of $YBCO_7$ are given in Table S 3, and some key stretching SFCs in the undistorted configuration decrease (c.f., Figure S 22b), reducing its stability.*

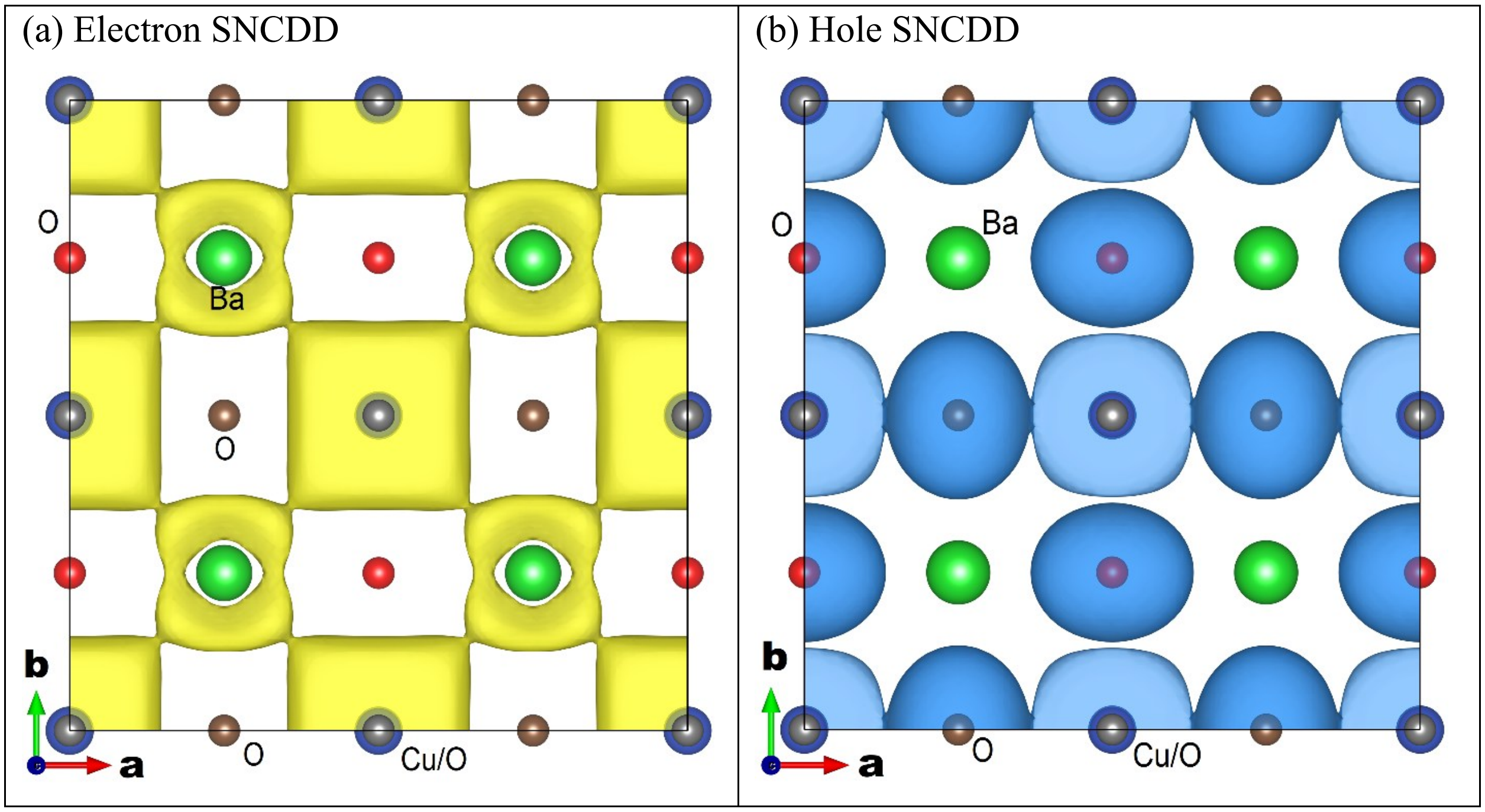

*Figure 5. Partial electron (a, in yellow) and hole (b, in blue) SNCDDs of $YBCO_7$ predicted by PBE, viewed along c-axis, showing the double 2D tunnels in (a) parallel to the a-b plane and SODTs in (b) along a-axis.*

# Supplementary Material

**Revealing Symmetry-Broken Superconducting Configurations by Density Functional Theory**

Zi-Kui Liu and Shun-Li Shang*

Department of Materials Science and Engineering, The Pennsylvania State University, University Park, PA 16802, USA
E-mail: sus26@psu.edu

## 1. Four Supplementary Tables

Table S 1. Examples of eight VASP-based configurations (POSCAR files) for pure elements in A1 (fcc), A4 (diamond), A6, and A7 crystal structures, as well as $YBCO_6$, $YBCO_7$, and $MgB_2$. The corresponding files, named as listed in this table, are provided as supplementary materials.

| Structure | Supercell | Examined pure elements or YBCO | Undistorted normal conducting configurations (NCCs) | Symmetry-broken superconducting configurations (SCCs) without relaxations |
|---|---|---|---|---|
| A1 (fcc) | 32-atom 2×2×2 | Al, Pb, Cu, Ag, Au, Rh, Ir, Pd, Pt, Ca, and Sr | POS1_FCC.txt | POS2_FCC.txt |
| A4 (diamond) | 64-atom 2×2×2 | Si, Ge, and Sn | POS3_A4.txt | POS4_A4.txt |
| A6 | 36-atom 2×2×2 | In | POS5_A6.txt | POS6_A6.txt |
| A7 | 54-atom 2×2×2 | As, Sb, and Bi | POS7_A7.txt | POS8_A7.txt |
| P4/mmm (#123) | 48-atom 2×2×1 | $YBCO_6$ | POS9_YBCO6.txt (NM case) | POS10_YBCO6.txt (NM case) |
| Pmmm (#47) | 52-atom 2×2×1 | $YBCO_7$ | POS11_YBCO7.txt (NM case) | POS12_YBCO7.txt (NM case) |
| P6/mmm | 24-atom 2×2×2 | $MgB_2$ | POS13_MgB2.txt | POS14_MgB2.txt |

Table S 2. Settings to generate SCCs, to perform DFT-based calculations, and to plot electron SNCDDs, including the structures (str.), the initial $\Delta d_{ini}$ to perturb atoms (cf., Eq. 4), the X-C functionals, the k-point meshes, cutoff energy ($E_{cut}$ in eV, determined by the VASP setting of "PREC = High" for pure elements), the minimum (Fmin) and the maximum (Fmax) charge density difference and the levels to plot electron SNCDDs ($F_{level}$) with charge gain.

| Elem | Str. | $\Delta d_{ini}$ | X-C | *k*-mesh | $E_{cut}$ | $F_{min}$ | $F_{max}$ | $F_{level}$ | Figure |
|---|---|---|---|---|---|---|---|---|---|
| Al | fcc | 0.5 | PBE | 9×9×9 | 312.4 | -0.0067922 | 0.00683086 | 0.00091 | Figure 1 |
| (3) [a] | | 0.5 | r$^2$SCAN | | | -0.00655679 | 0.00649004 | 0.0006 | Figure S 2 |
| Pb_d | fcc | 0.5 | PBE | 9×9×9 | 309.2 | -0.251745 | 0.250753 | 0.00144 | Figure 2 |
| (14) [a] | | 0.7 | r$^2$SCAN | | | -0.25938 | 0.258411 | 0.00142 | Figure S 6 |
| Cu | fcc | 0.5 | PBE | 9×9×9 | 384.1 | -0.550911 | 0.551039 | 0.00228 | Figure S 7 |
| (11) [a] | | 0.5 | r$^2$SCAN | | | -0.540195 | 0.547063 | 0.0025 | |
| Ag | fcc | 0.7 | PBE | 9×9×9 | 324.8 | -0.204061 | 0.205891 | 0.00215 | Figure S 8 |
| (11) [a] | | 0.7 | r$^2$SCAN | | | -0.202644 | 0.205472 | 0.00214 | |
| Au | fcc | 0.7 | PBE | 9×9×9 | 298.9 | -0.101173 | 0.100657 | 0.003 | Figure S 9 |
| (11) [a] | | 0.7 | r$^2$SCAN | | | -0.100973 | 0.100535 | 0.0032 | |
| Rh_pv | fcc | 0.1 | PBE | 9×9×9 | 321.6 | -0.0313492 | 0.03137 | 0.000375 | Figure S 10 |
| (15) [a] | | 0.1 | r$^2$SCAN | | | -0.0312419 | 0.0313652 | 0.000385 | |
| Ir | fcc | 0.5 | PBE | 9×9×9 | 274.1 | -0.0432119 | 0.0442711 | 0.00508 | Figure S 11 |
| (9) [a] | | 0.5 | r$^2$SCAN | | | -0.0437308 | 0.0445784 | 0.00528 | |
| Pd | fcc | 0.5 | PBE | 9×9×9 | 326.2 | -0.150707 | 0.150854 | 0.0031 | Figure S 12 |
| (10) [a] | | 0.5 | r$^2$SCAN | | | -0.12099 | 0.121396 | 0.00237 | |
| Pt | fcc | 0.5 | PBE | 9×9×9 | 299.4 | -0.0561032 | 0.0560233 | 0.00325 | Figure S 13 |
| (10) [a] | | 0.5 | r$^2$SCAN | | | -0.0685293 | 0.0694616 | 0.0046 | |
| Ca_sv | fcc | 0.5 | PBE | 9×9×9 | 346.6 | -0.213748 | 0.213162 | 0.000043 | Figure S 14 |
| (10) [a] | | 0.5 | r$^2$SCAN | 7×7×7 | | -0.210528 | 0.212086 | 0.000034 | |
| Sr_sv | fcc | 0.5 | PBE | 9×9×9 | 298.2 | -0.109094 | 0.108409 | 5.10E-05 | Figure S 15 |
| (10) [a] | | 0.1 | r$^2$SCAN | | | -0.0153178 | 0.0152335 | 3.50E-06 | |
| Si | A4 | 0.3 | PBE | 7×7×7 | 318.9 | -0.00760139 | 0.00755245 | 0.00167 | Figure S 16 |
| (4) [a] | | 0.5 | r$^2$SCAN | | | -0.00645129 | 0.00642902 | 0.00165 | |
| Ge_d | A4 | 0.5 | PBE | 7×7×7 | 403.4 | -0.0943039 | 0.0950664 | 0.00065 | Figure S 17 |
| (14) [a] | | 0.3 | r$^2$SCAN | | | -0.327742 | 0.33064 | 0.00265 | |
| Sn_d | A4 | 0.5 | PBE | 7×7×7 | 313.4 | -0.123664 | 0.12295 | 0.00176 | Figure S 18 |
| (14) [a] | | 0.3 | r$^2$SCAN | 5×5×5 | | -0.148192 | 0.1495 | 0.0021 | |
| In_d | A6 | 0.4 | PBE | 9×9×9 | 311.0 | -0.352342 | 0.281976 | 0.00206 | Figure S 19 |
| (13) [a] | | 0.4 | r$^2$SCAN | 9×9×9 | | -0.281243 | 0.278783 | 0.00108 | |
| As | A7 | 0.4 | PBE | 9×9×9 | 271.3 | -0.00740275 | 0.00745658 | 0.00154 | Figure 3 |
| (5) [a] | | 0.4 | r$^2$SCAN | 9×9×9 | | -0.00655503 | 0.00660198 | 0.00071 | |
| Sb | A7 | 0.4 | PBE | 9×9×9 | 223.7 | -0.00647086 | 0.00657628 | 0.000463 | Figure S 20 |
| (5) [a] | | 0.6 | r$^2$SCAN | 9×9×9 | | -0.0103783 | 0.0103252 | 0.00148 | |
| Bi | A7 | 0.4 | PBE | 9×9×9 | 136.5 | -0.0100346 | 0.0101007 | 0.00155 | Figure S 21 |
| (5) [a] | | 0.6 | r$^2$SCAN | 9×9×9 | | -0.00958677 | 0.0095138 | 0.0017 | |
| $YBCO_6$ | | | PBE | 35 [b] | 520 | -1.36947 | 1.37642 | 0.00014 | Figure S 23 |
| | | | r$^2$SCAN | 35 [b] | | -1.43964 | 1.46245 | 0.00014 | Figure S 24 |
| $YBCO_7$ | | | PBE | 35 [b] | 520 | -2.43893 | 2.47636 | 0.0065 | Figure 5<br>Figure S 25 |
| | | | r$^2$SCAN | 35 [b] | | -2.26760 | 2.28300 | 0.000415 | Figure S 26 |
| $MgB_2$ | | 0.06[c] | PBE | 7×7×7 | 504.9 | -0.263448 | 0.266759 | 0.00129 | Figure S 29 |
| | | | r$^2$SCAN | 7×7×7 | | -0.216600 | 0.223696 | 0.000145 | |

[a] Valence electrons used in the present first-principles calculations. In addition, the suffixes "_sv, _pv, or _d" after the symbols of some elements indicate the s, p, and d states are considered as valence states.
[b] Length used to determine the subdivisions of *k*-point meshes.
[c] Value used as direct (fractional) coordinates.

Table S 3. Crystallographic details of $YBCO_7$ and $YBCO_6$ by experiments *[1,2]* and by DFT-based predictions (showing in the parentheses by PBE and listing only the different values), including lattice parameters $a$, $b$, and $c$ (in Å) and atomic positions $x$, $y$, and $z$. Note that Figure 4(a) illustrates the relaxed configuration of $YBCO_7$ and Figure S 22(a) plots the undistorted configuration of $YBCO_7$.

| Atoms | $YBCO_7$ with space group $Pmmm$ | | | $YBCO_6$ with space group $P4/mmm$ | | |
|---|---|---|---|---|---|---|
| | x or a | y or b | z or c | x or a | y or b | z or c |
| | 3.820 (3.837) | 3.886 (3.919) | 11.684 (11.869) | 3.859 (3.857) | 3.859 (3.857) | 11.814 (11.936) |
| Ba | 0.5 | 0.5 | 0.1839 (0.1805) | 0.5 | 0.5 | 0.1946 (0.1939) |
| Y | 0.5 | 0.5 | 0.5 | 0.5 | 0.5 | 0.5 |
| Cu1 | 0 | 0 | 0 | 0 | 0 | 0 |
| Cu2 | 0 | 0 | 0.3550 (0.3543) | 0 | 0 | 0.3611 (0.3647) |
| O1* | 0 | 0.5 | 0 | | | |
| O2 | 0.5 | 0 | 0.3782 (0.3807) | 0 | 0.5 | 0.3798 (0.3801) |
| O3 | 0 | 0.5 | 0.3769 (0.3792) | | | |
| O4 | 0 | 0 | 0.1584 (0.1581) | 0 | 0 | 0.1524 (0.1513) |

*Occupancy of 0.910 by experiments *[1]*.

Table S 4. Equilibrium properties of $YBCO_6$ and $YBCO_7$ by the present DFT calculations and energy-volume EOS using PBE for NM configurations and r$^2$SCAN for AFM calculations in comparison with available experimental data in the literature. The absolute values of the projected magnetic moment (MM) for AFM Cu by r$^2$SCAN are also reported, while the MM values for other atoms including non-AFM Cu are close to zero (< 0.02 $\mu_B$/atom). Note that the EOS fitted $V_0$ values are slightly larger than the fully relaxed $V_0$ values by about 0.55% for $YBCO_7$ and 0.73% for $YBCO_6$.

| Materials | $V_0$ (Å$^3$/f.u.) | $B_0$ (GPa) | $B'$ | MM ($\mu_B$/atom) | Notes |
|---|---|---|---|---|---|
| $YBCO_7$ | 173.88 | 115.5 | 7.18 | 0.42 | Calc. (This work, r$^2$SCAN) |
| | 179.05 | 101.6 | 6.65 | | Calc. (This work, PBE) |
| | 173.43 [a] | 115 [b] | | | Expt. |
| $YBCO_6$ | 178.03 | 95.3 | 4.87 | 0.49 | Calc. (This work, r$^2$SCAN) |
| | 183.7 | 81.1 | 5.56 | | Calc. (This work, PBE) |
| | 176.0 [c] | | | | Expt. |

[a] Measured data at 297 K by X-ray and neutron powder diffraction [1].

[b] This value was believed as the best bulk modulus using high-pressure X-ray diffraction [3].

[c] Measured data of single crystal X-ray diffraction [2].

# 1 Supplementary Figure S 1: Test of Al

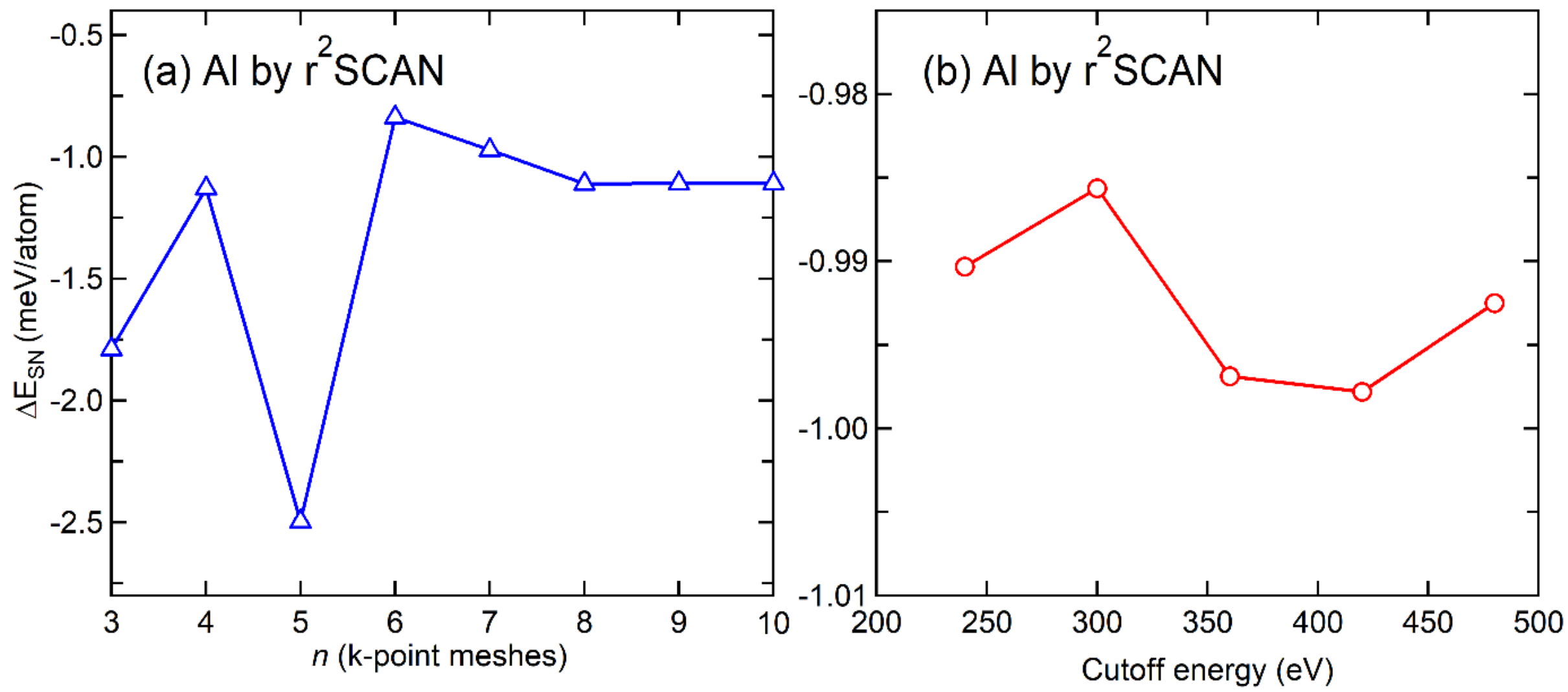

Figure S 1. Convergence tests of the predicted energy difference, $\Delta E_{SN}$, between SCC and NCC for Al using $r^2$SCAN: (a) k-point meshes of n×n×n and (b) cutoff energy.

## 2 Supplementary figures of Al, Pb, Cu

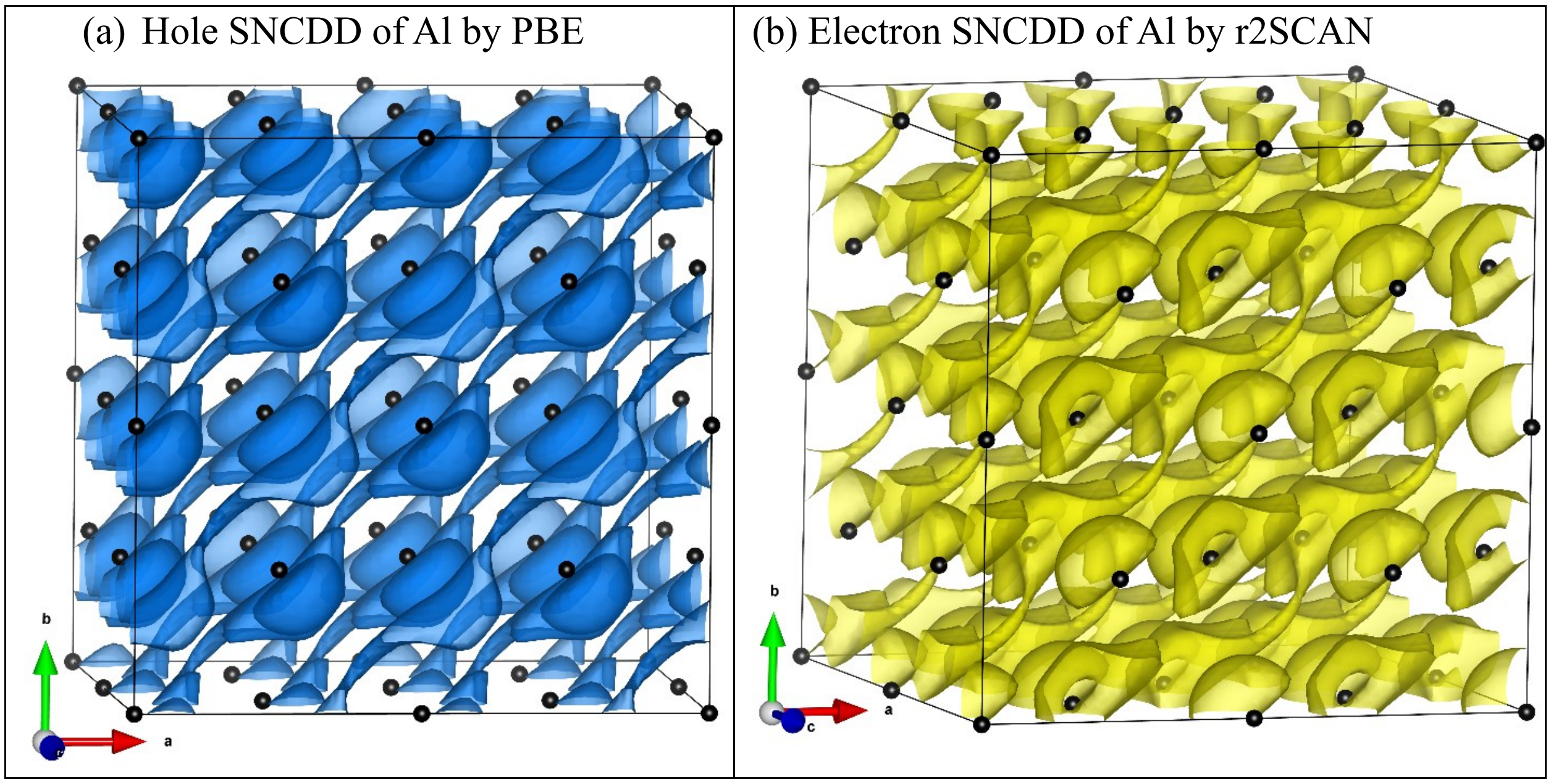

Figure S 2. SNCDD of Al. PBE predicted hole SNCDD due to charge loss (a, in blue) and r2SCAN predicted electron SNCDD due to charge gain (b, in yellow); indicating the formation of SODTs along [110] direction.

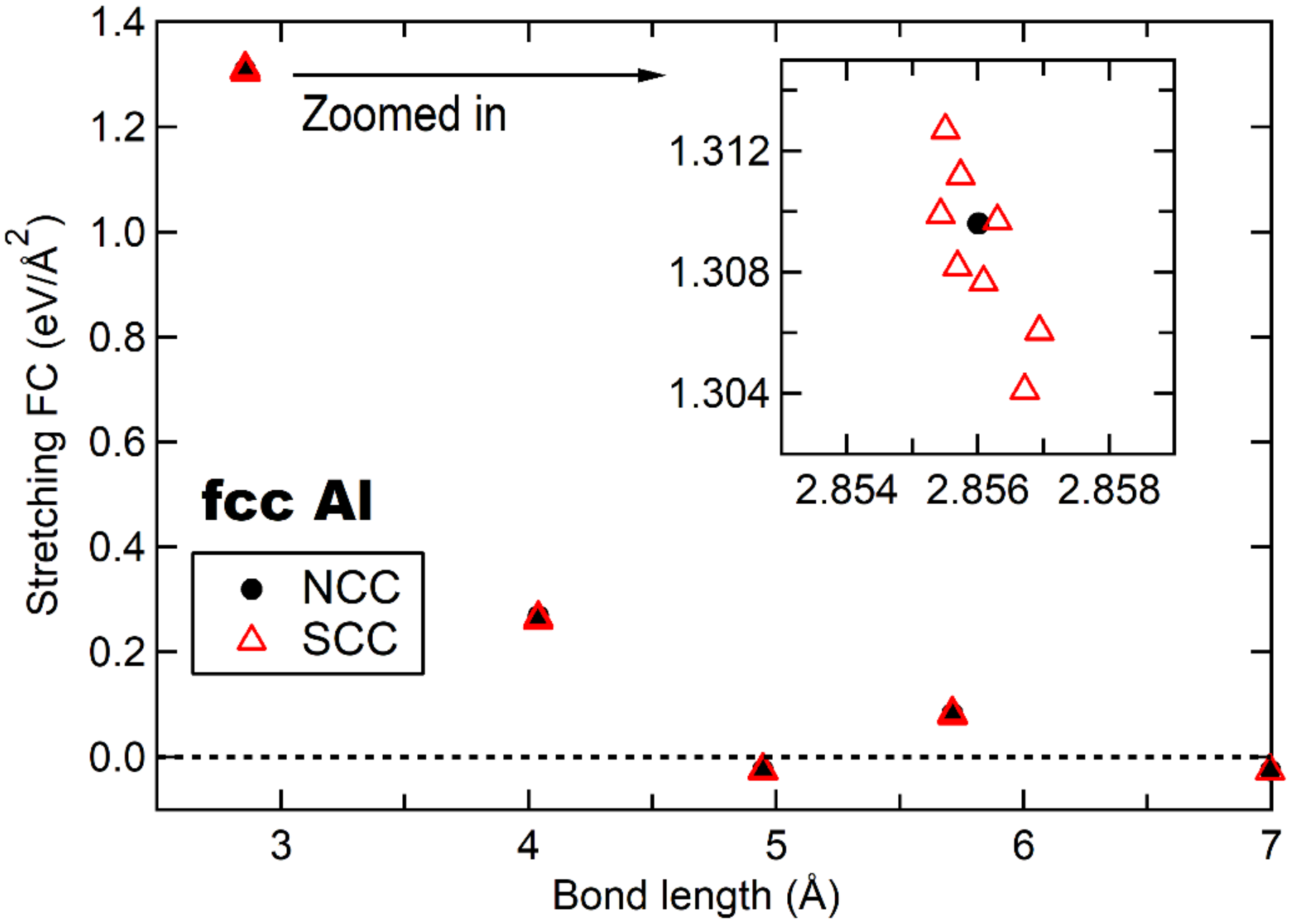

*Figure S 3. Stretching force constants (SFC's) as a function of bond length for fcc Al predicted by PBE using the 32-atom NCC and SCC by phonon calculations, respectively.*

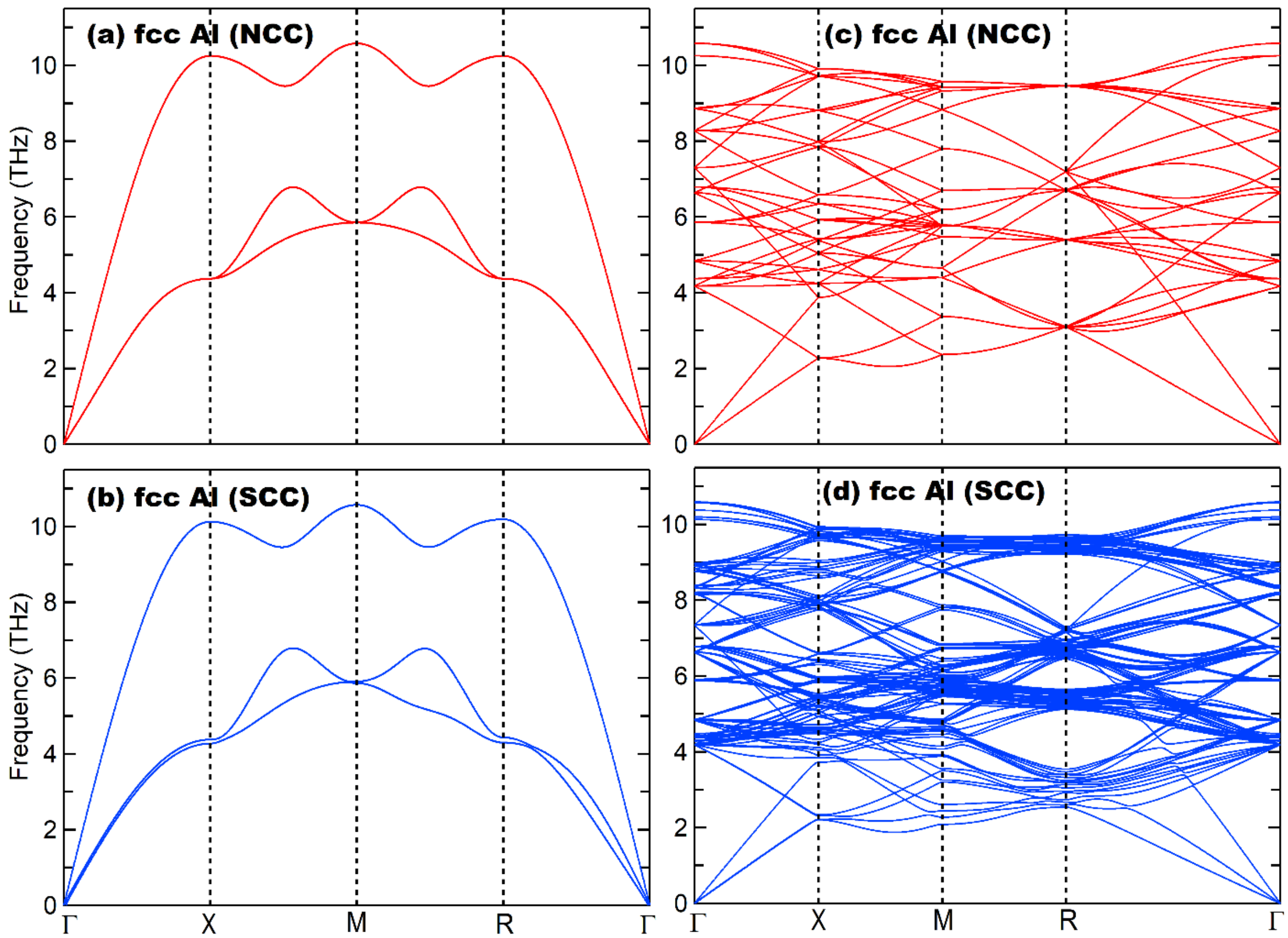

*Figure S 4. Phonon dispersions of fcc Al (a and b) along the high-symmetry directions based on the 1-atom primitive cells for NCC and SCC, respectively, by PBE; and (c and d) along high-symmetry directions based on the 32-atom supercells for NCC and SCC, respectively, by PBE.*

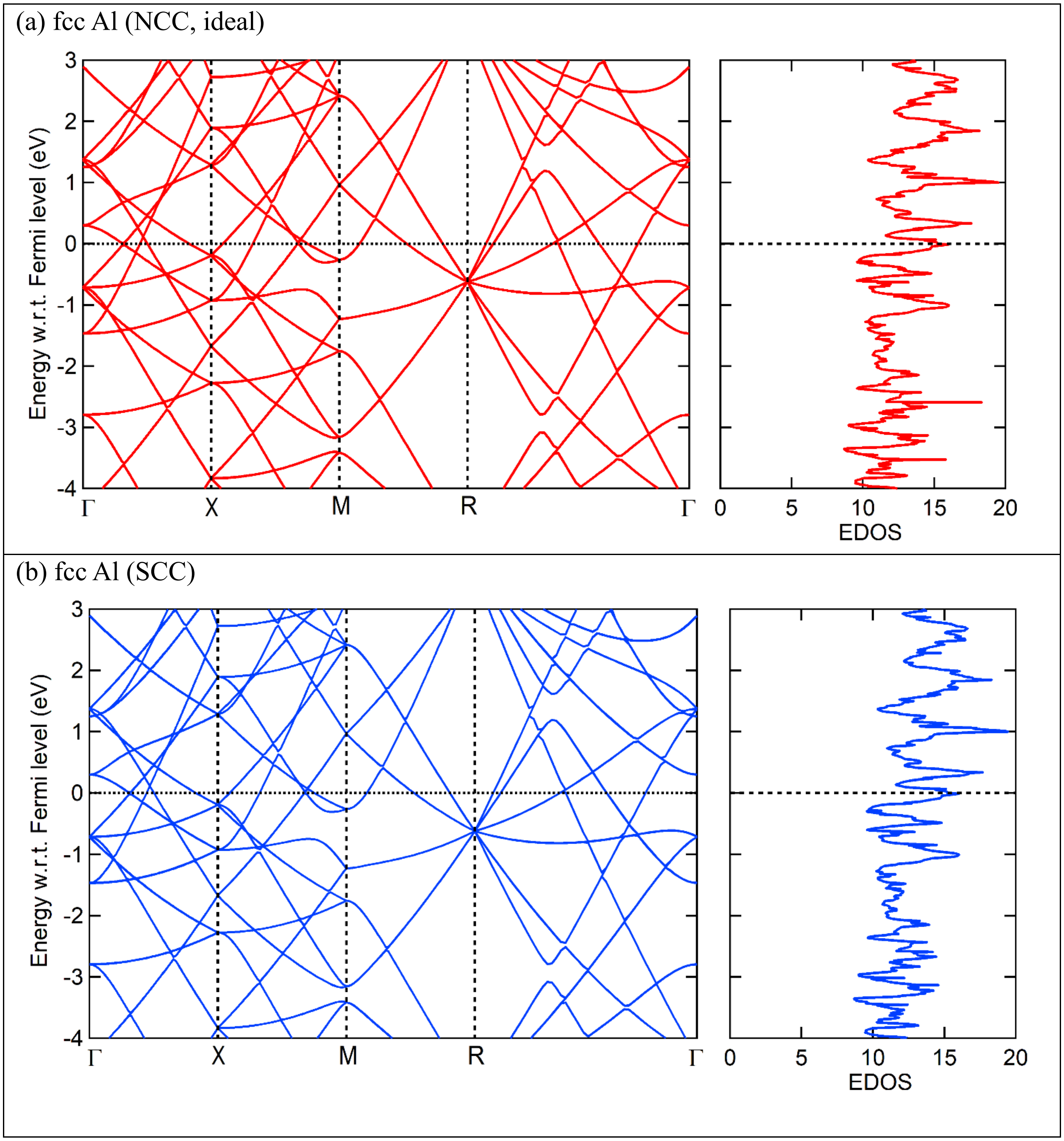

*Figure S 5. Band structures and electron density of states (eDOS) of fcc Al for (a) NCC and (b) SCC, predicted by PBE.*

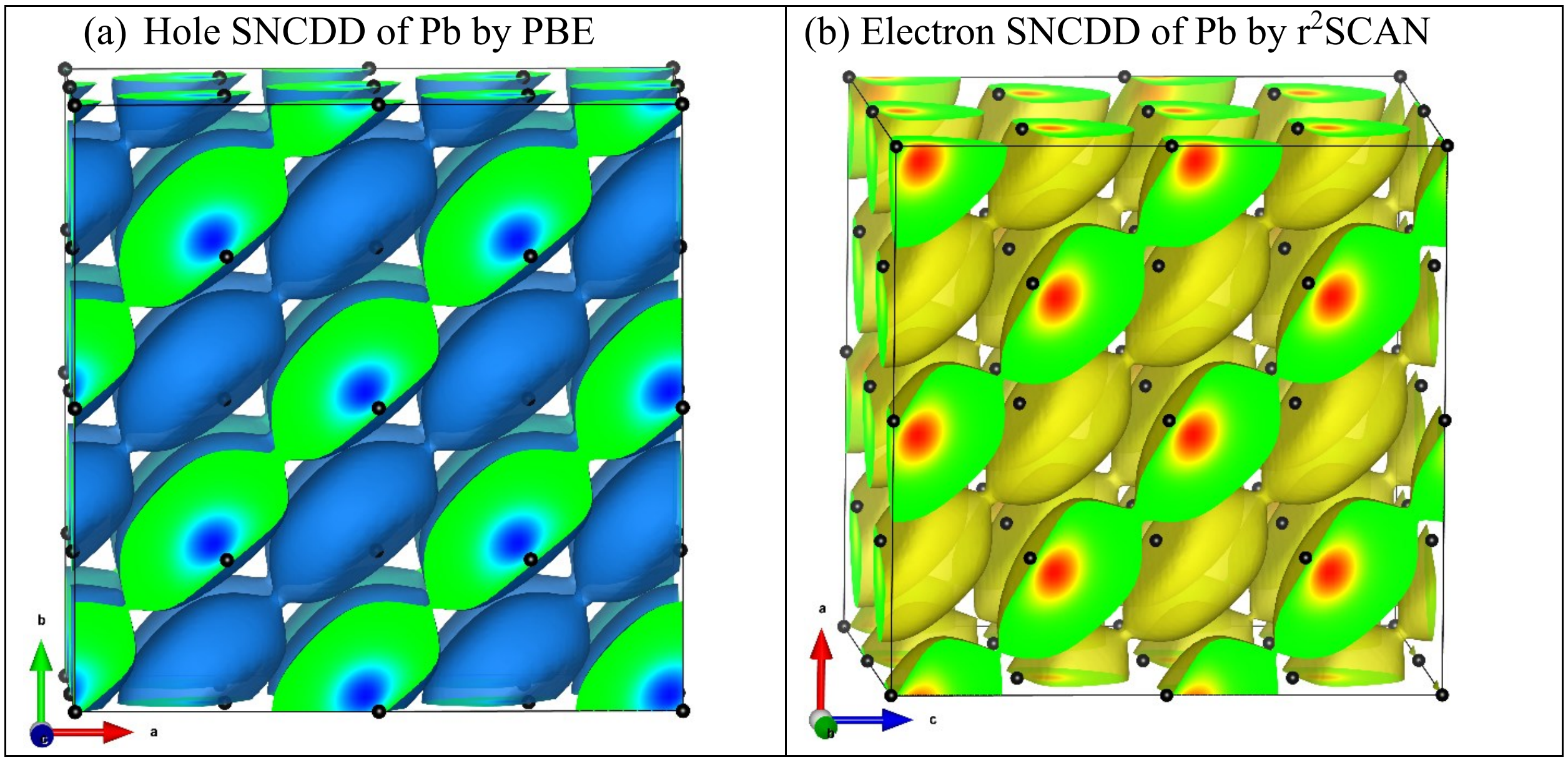

Figure S 6. SNCDD of Pb. PBE predicted hole SNCDD due to charge loss (a, in blue) and $r^2$SCAN predicted electron SNCDD due to charge gain (b, in yellow), indicating the formation of SODTs along [110] or [101] direction.

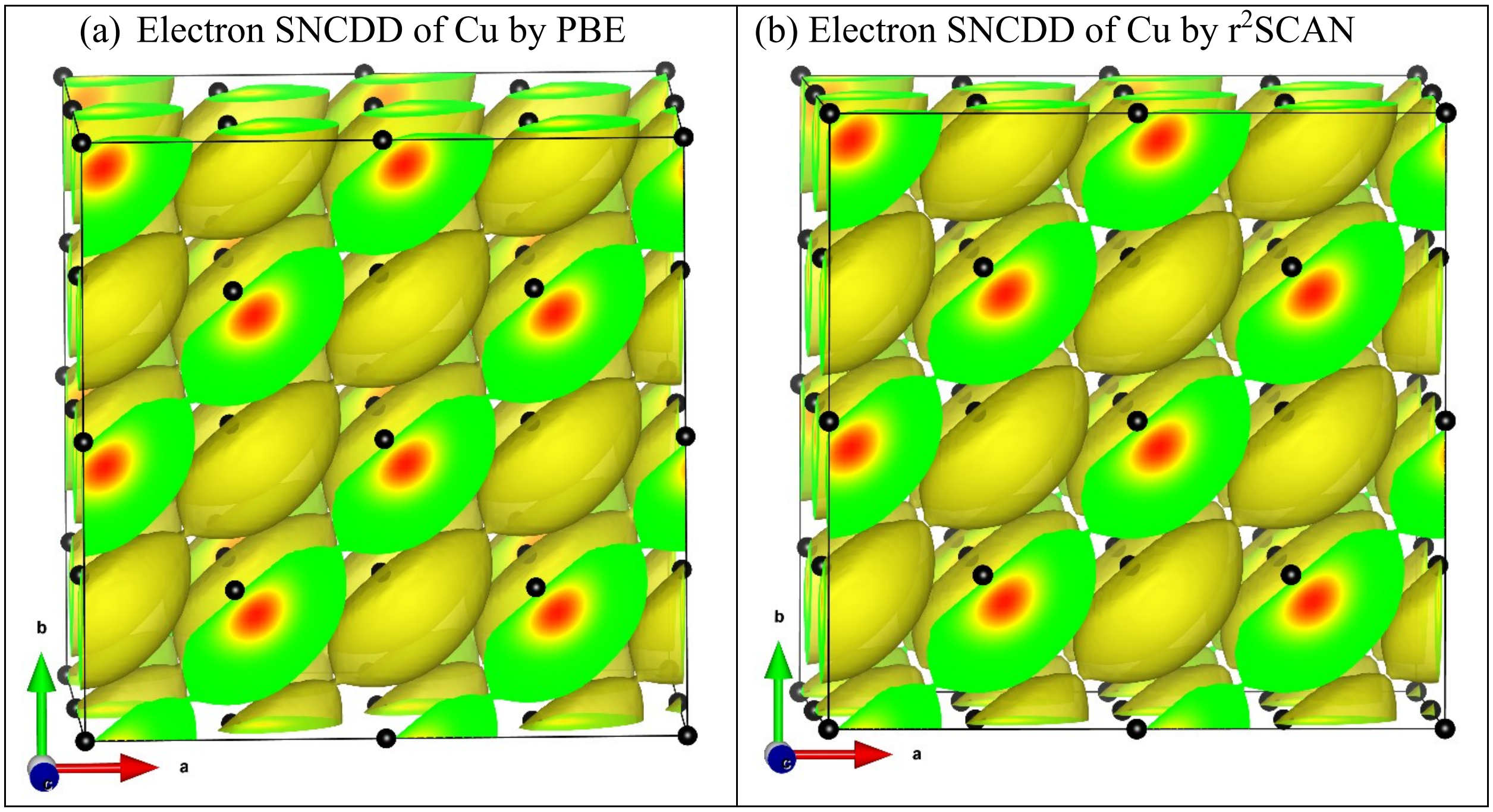

Figure S 7. SNCDD of Cu. PBE predicted electron SNCDD due to charge gain (a, in yellow) and $r^2$SCAN predicted electron SNCDD due to charge gain (b, in yellow), indicating the formation of SODTs along [110] direction.

## 3 Supplementary figures of Ag, Au

(a) Electron SNCDD of Ag by PBE | (b) Electron SNCDD of Ag by $r^2$SCAN

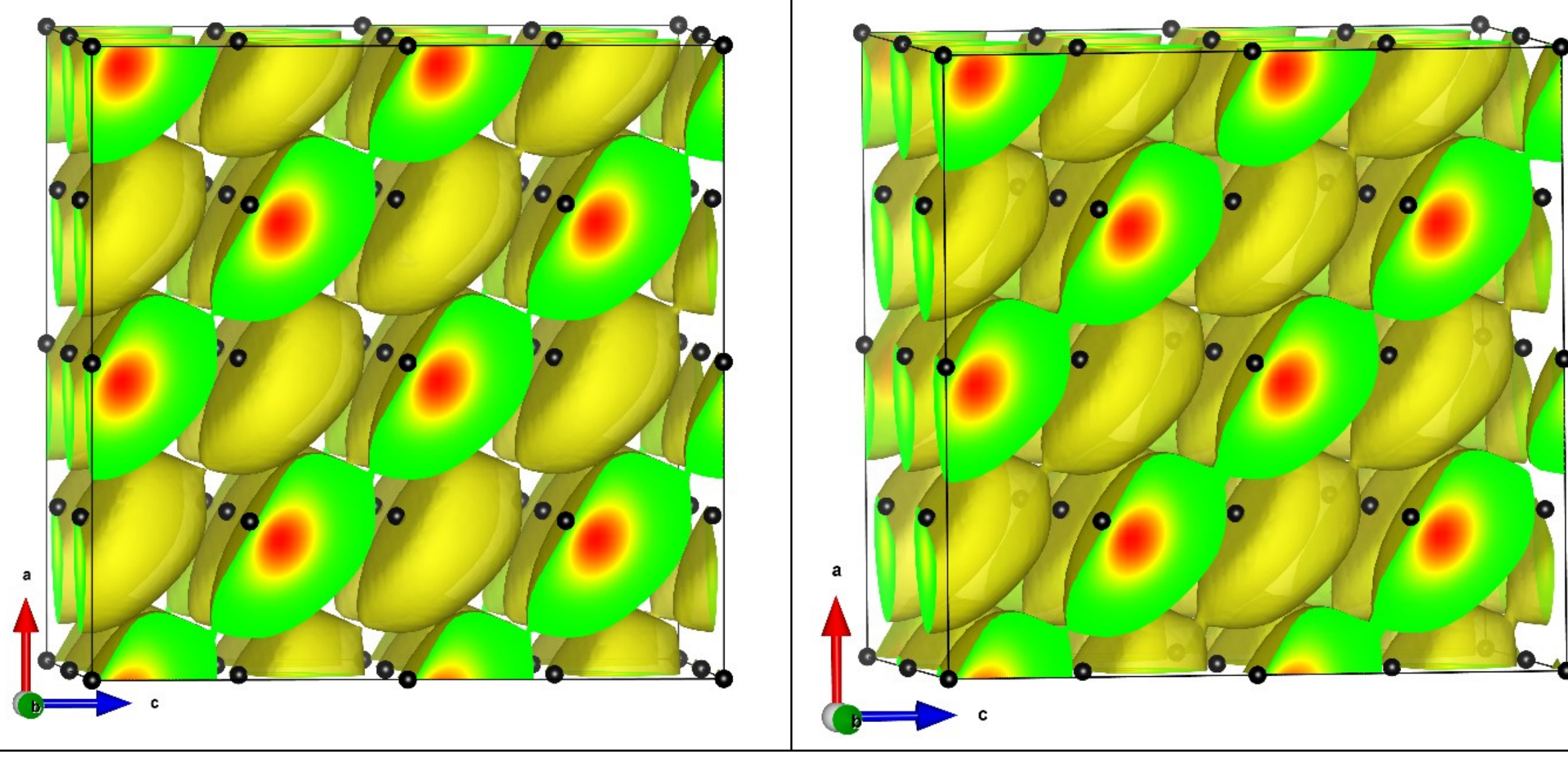

Figure S 8. SNCDD of Ag. PBE predicted electron SNCDD due to charge gain (a, in yellow), and $r^2$SCAN predicted electron SNCDD due to charge gain (b, in yellow), indicating the formation of SODTs along [101] direction.

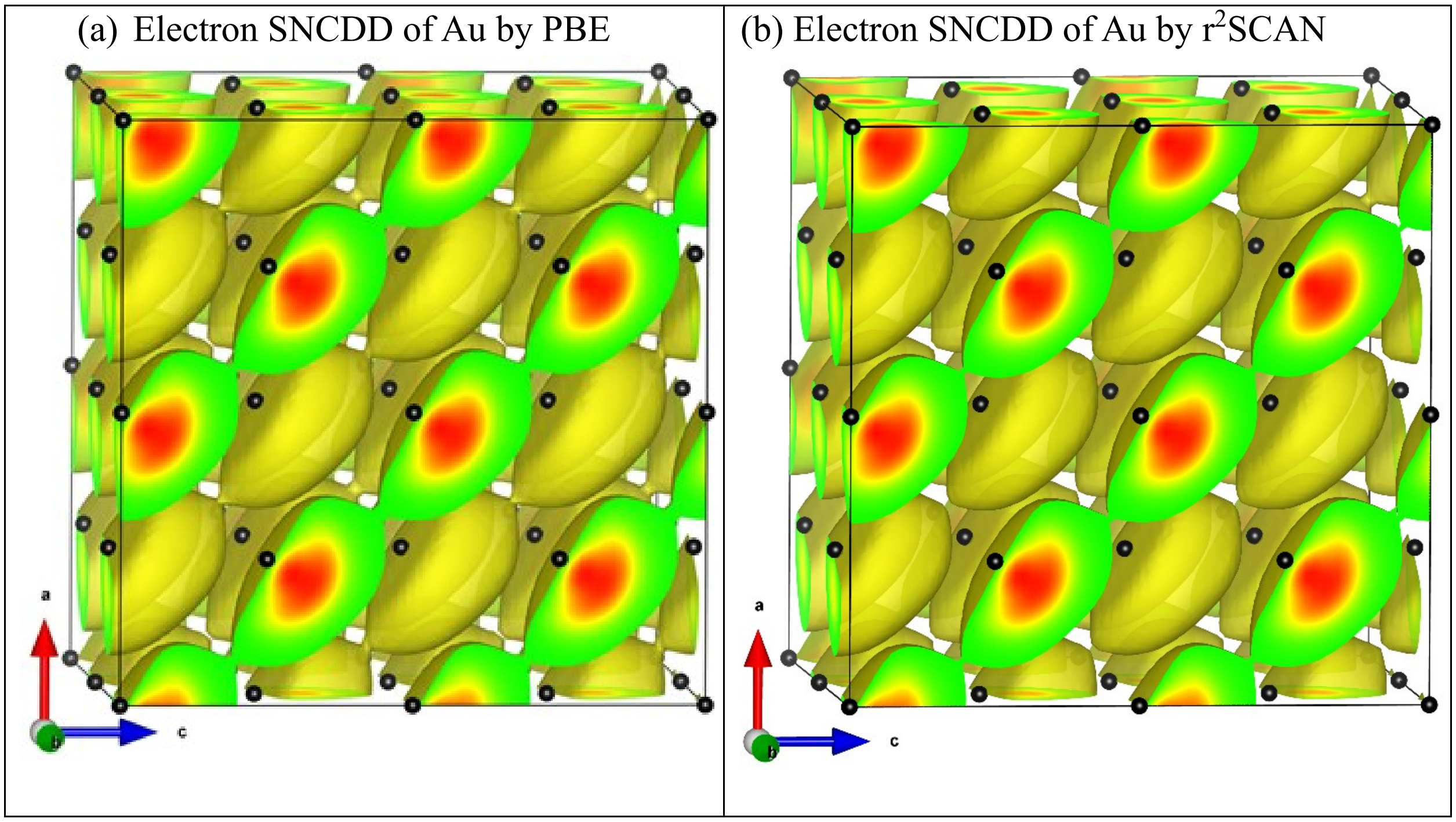

Figure S 9. SNCDD of Au. PBE predicted electron SNCDD due to charge gain (a, in yellow), and $r^2$SCAN predicted electron SNCDD due to charge gain (b, in yellow), indicating the formation of SODTs along [101] direction.

## 4 Supplementary figures of Rh, Ir

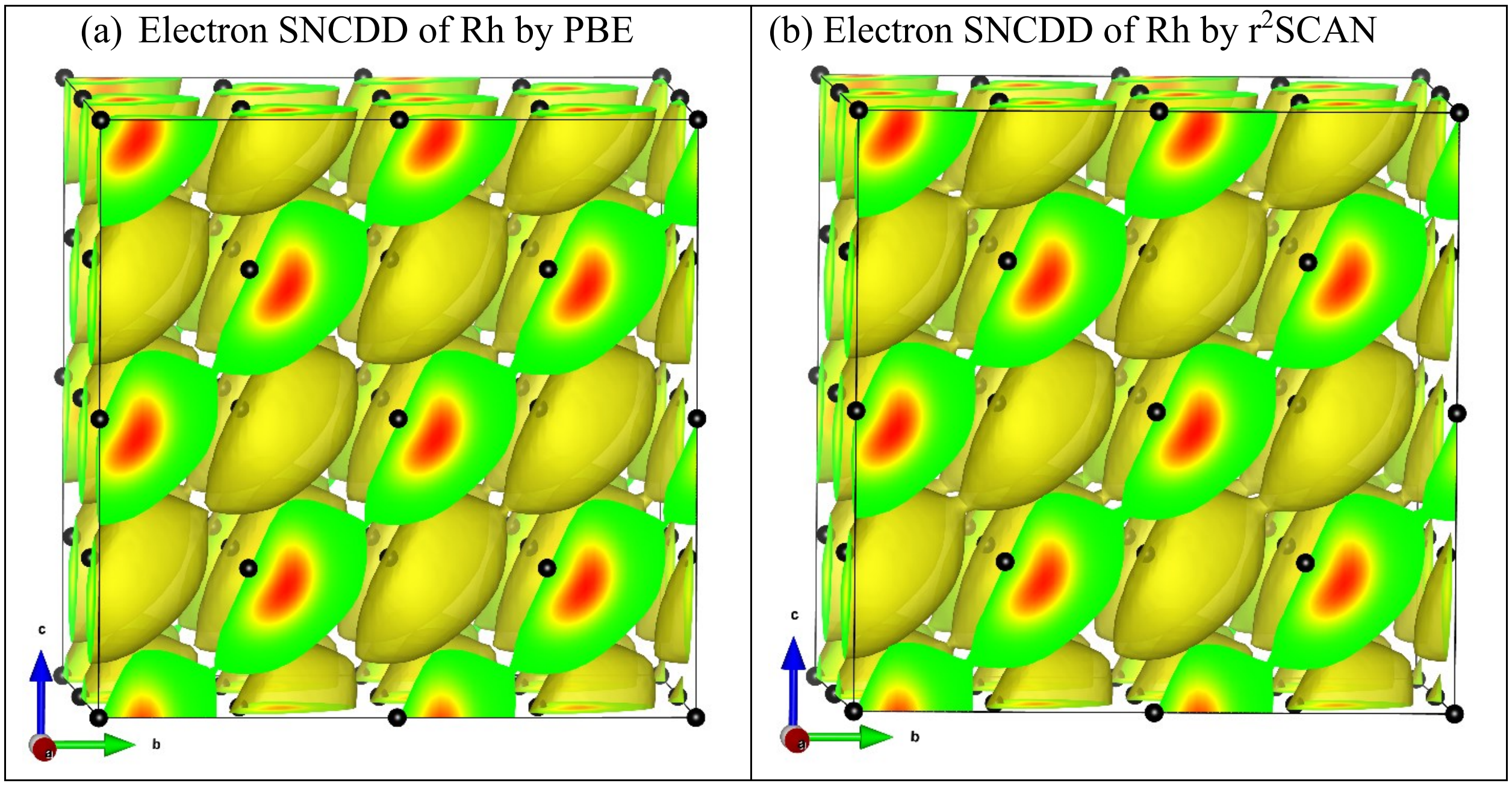

Figure S 10. SNCDD of Rh. PBE predicted electron SNCDD due to charge gain (a, in yellow), and r$^2$SCAN predicted electron SNCDD due to charge gain (b, in yellow), indicating the formation of SODTs along [011] direction.

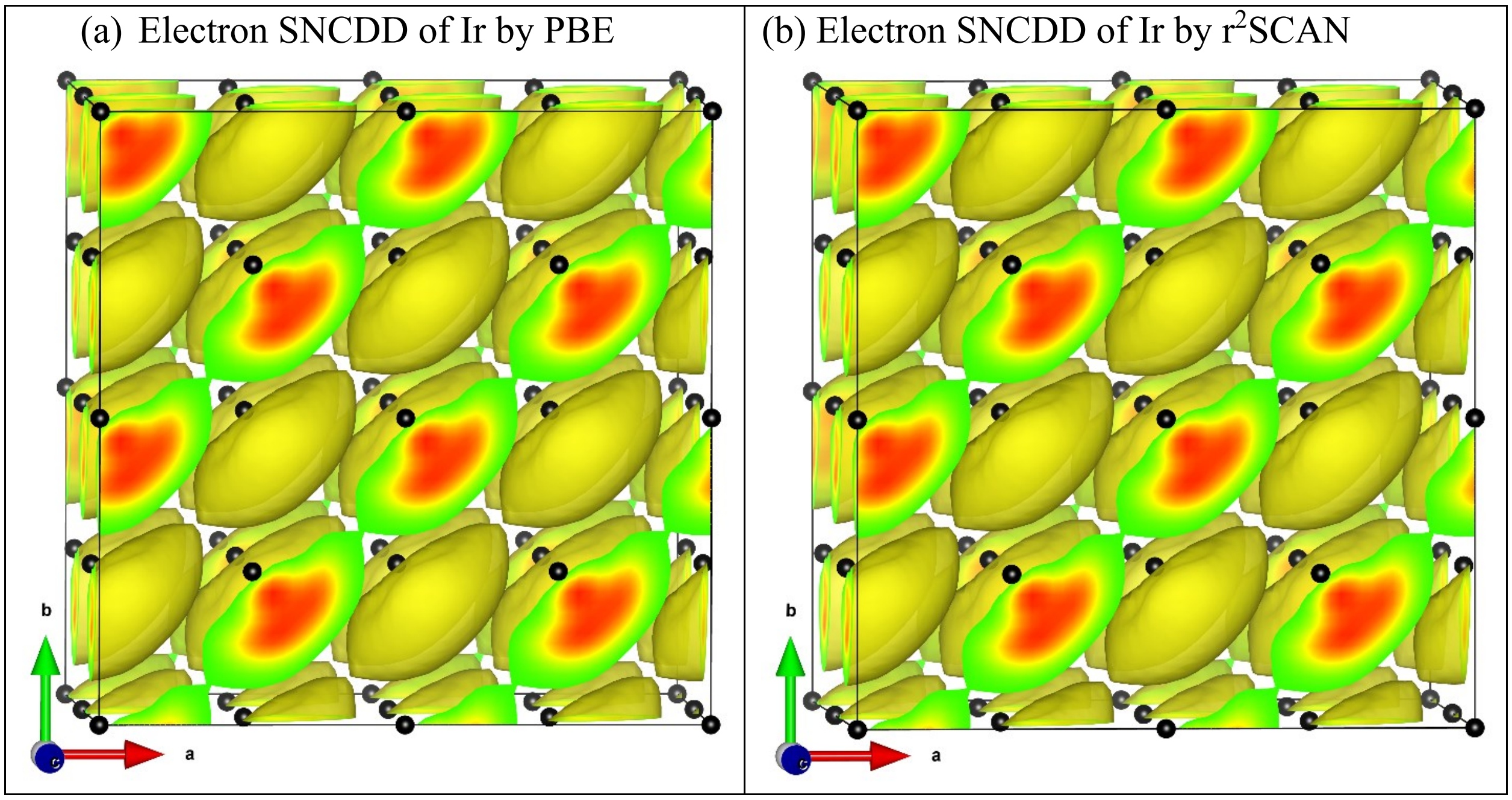

Figure S 11. SNCDD of Ir. PBE predicted electron SNCDD due to charge gain (a, in yellow), and $r^2$SCAN predicted electron SNCDD due to charge gain (b, in yellow), indicating the formation of SODTs along [110] direction.

## 5 Supplementary figures of Pd, Pt

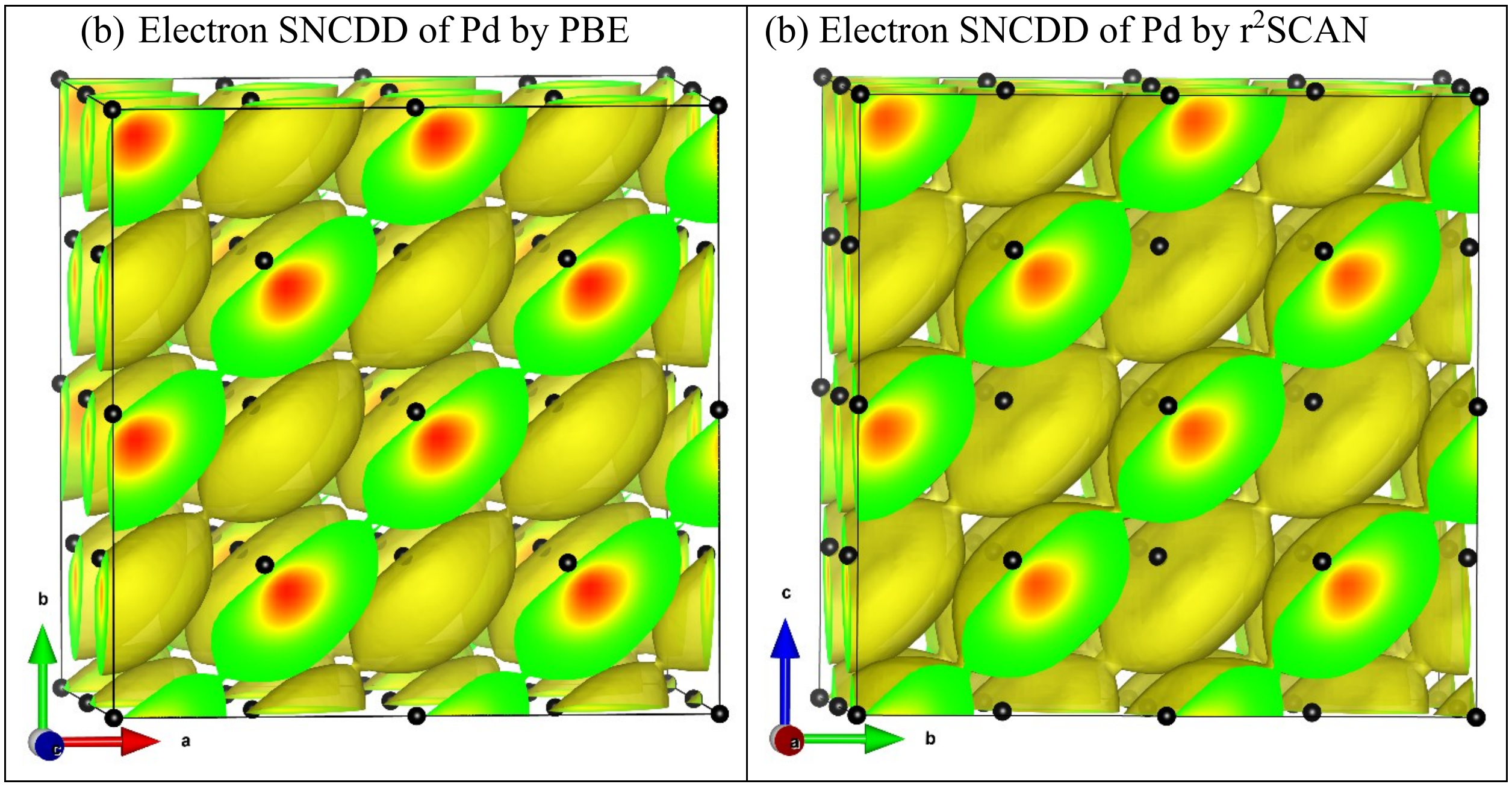

Figure S 12. SNCDD of Pd. PBE predicted electron SNCDD due to charge gain (a, in yellow), and $r^2$SCAN predicted electron SNCDD due to charge gain (b, in yellow), indicating the formation of SODTs along [110] or [011] direction.

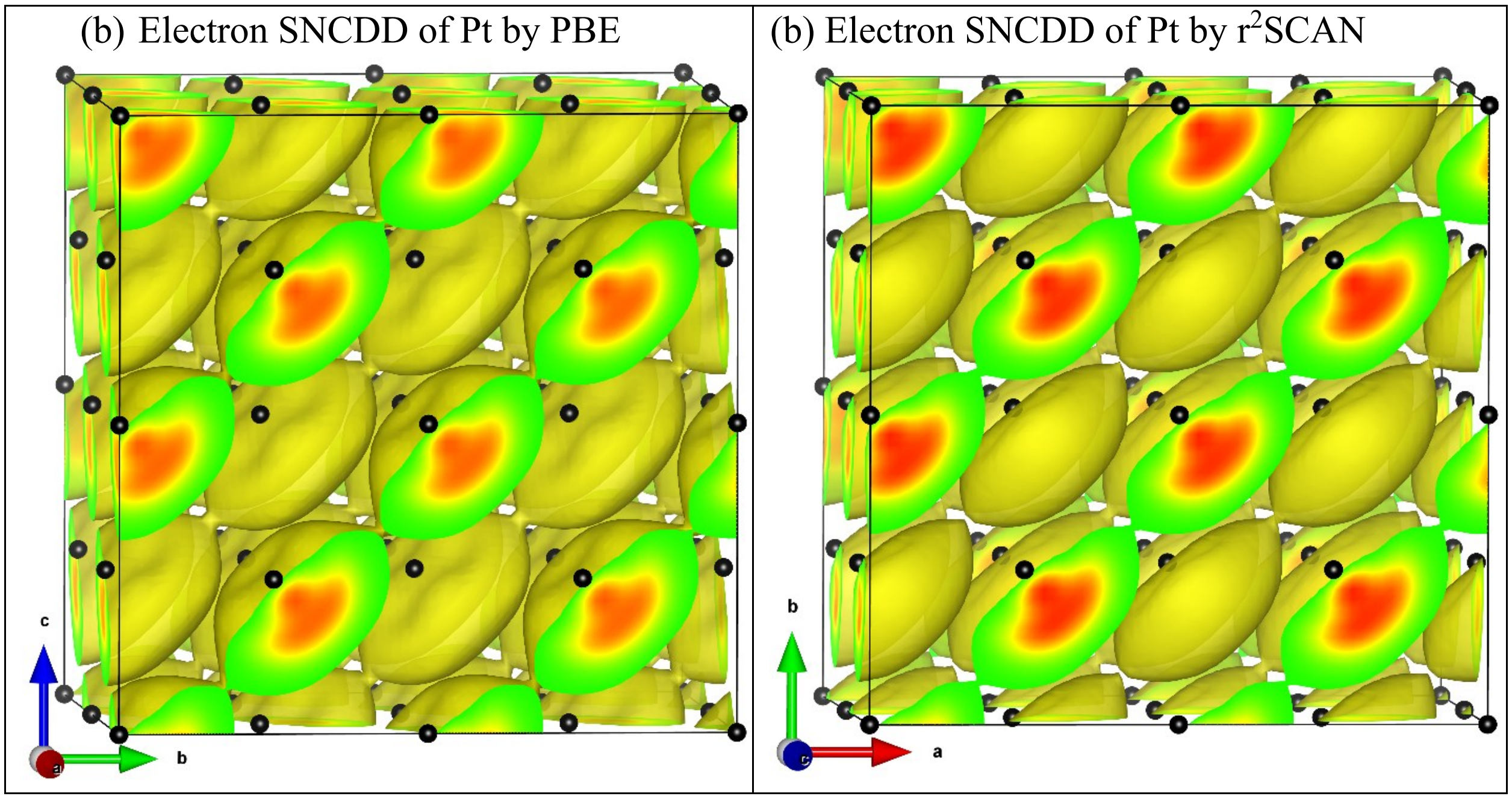

Figure S 13. SNCDD of Pt. PBE predicted electron SNCDD due to charge gain (a, in yellow), and $r^2$SCAN predicted electron SNCDD due to charge gain (b, in yellow), indicating the formation of SODTs along [011] or [110] direction.

## 6 Supplementary figures of Ca, Sr

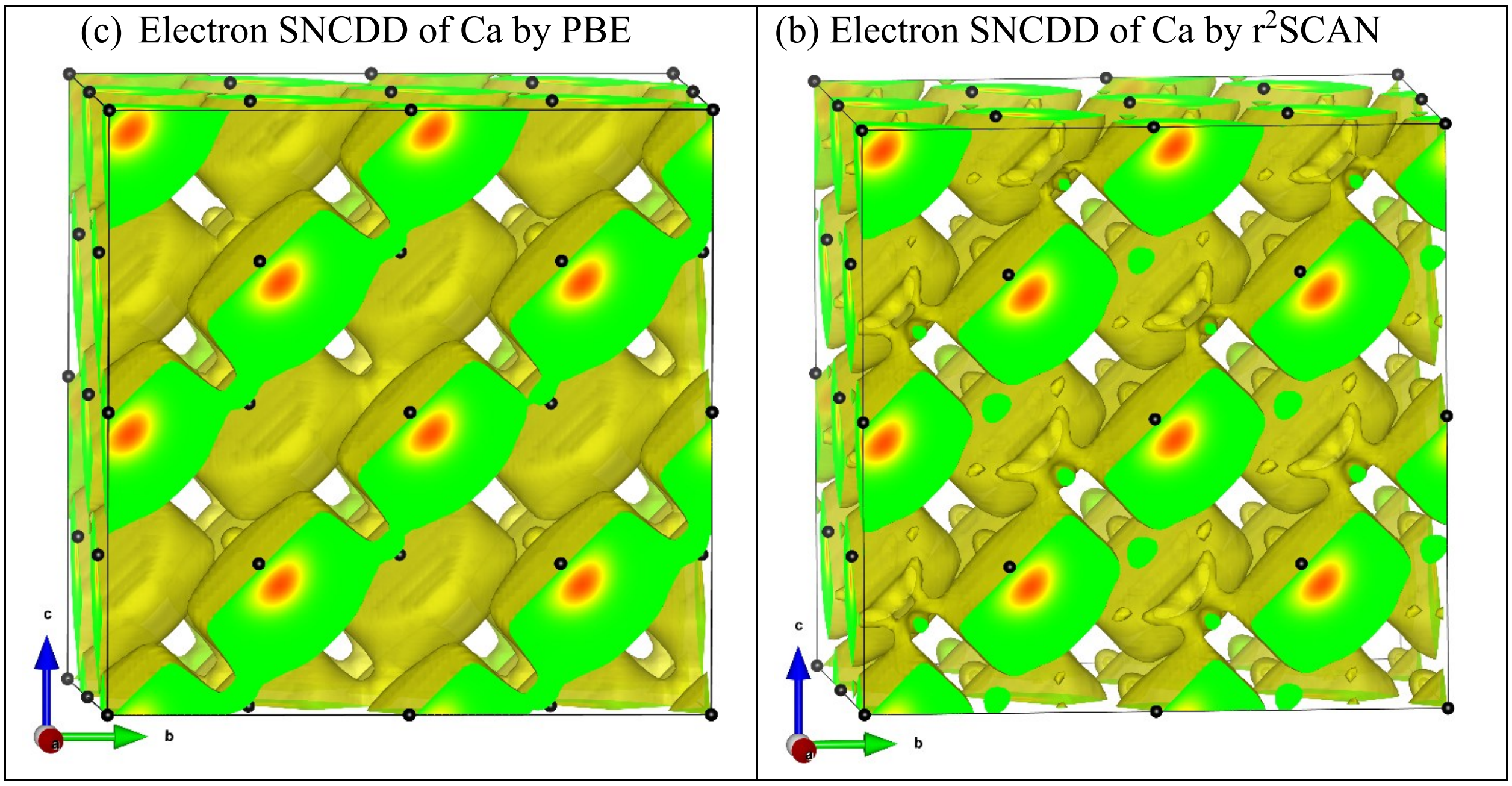

Figure S 14. SNCDD of Ca. PBE predicted electron SNCDD due to charge gain (a, in yellow), and r$^2$SCAN predicted electron SNCDD due to charge gain (b, in yellow), indicating the formation of SODTs along [011] by BPE but not by r$^2$SCAN. It would be beneficial to verify these results using additional X-C functionals, such an investigation is ongoing.

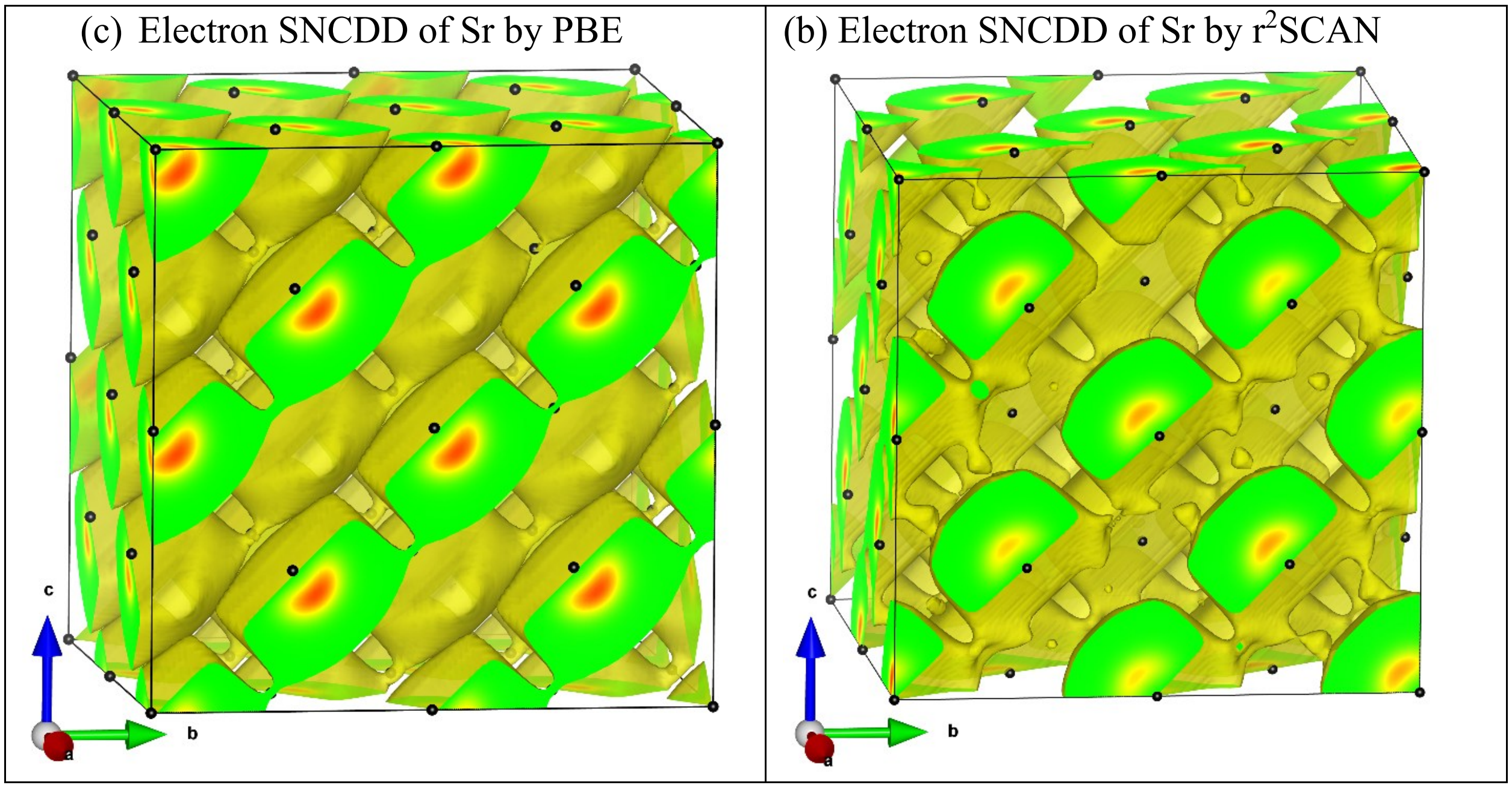

Figure S 15. SNCDD of Sr. PBE predicted electron SNCDD due to charge gain (a, in yellow), and $r^2$SCAN predicted electron SNCDD due to charge gain (b, in yellow), indicating the formation of SODTs along [011] by BPE but not by $r^2$SCAN. It would be beneficial to verify these results using additional X-C functionals, such an investigation is ongoing.

## 7 Supplementary figures of Si, Ge, Sn (A4)

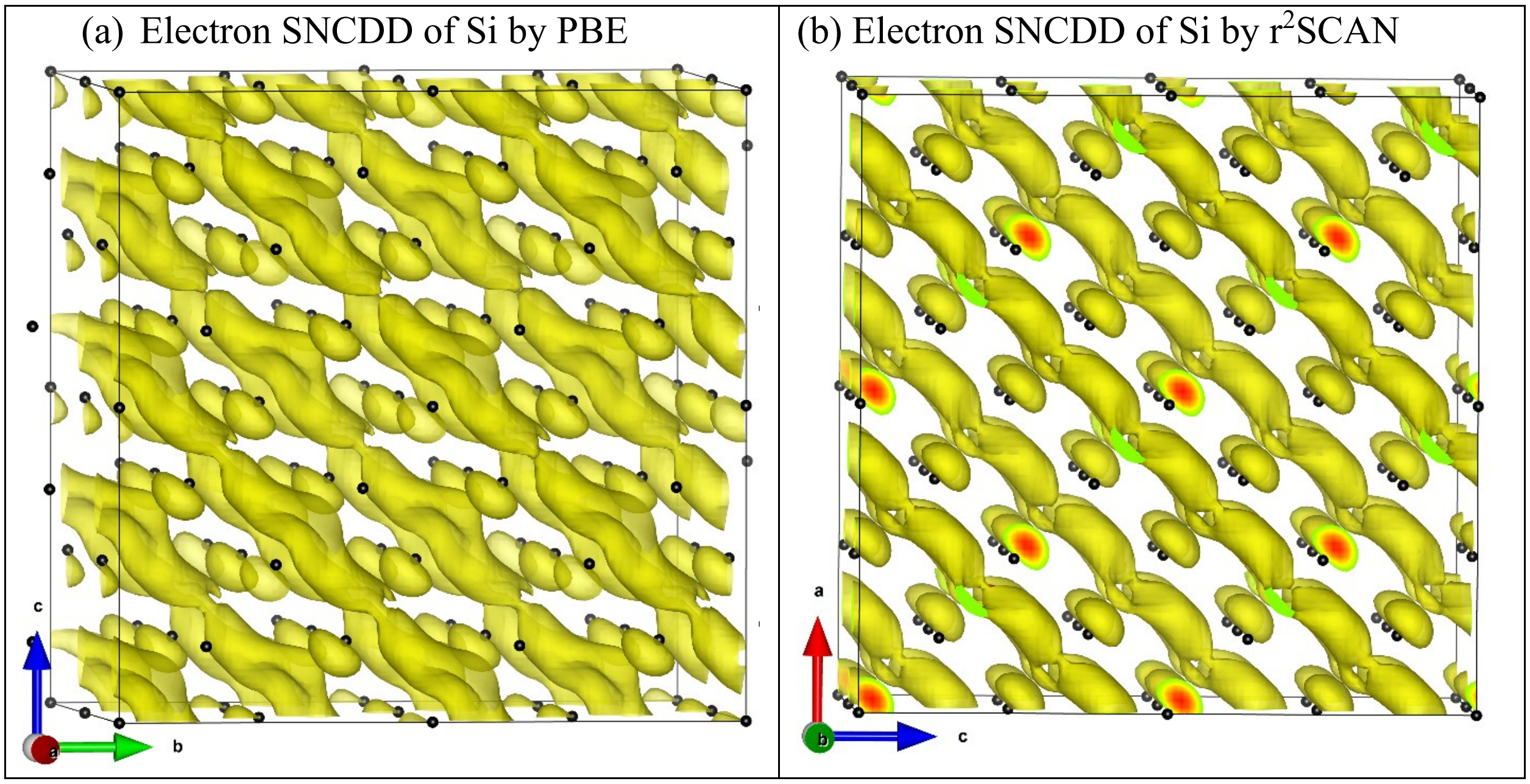

Figure S 16. SNCDD of Si. PBE predicted electron SNCDD due to charge gain (a, in yellow), and $r^2$SCAN predicted electron SNCDD due to charge gain (b, in yellow), indicating the formation of SODTs ($[0\bar{1}1]$ by PBE and $[10\bar{1}]$ by $r^2$SCAN). However, Si is semiconductor without free electrons and holes at the Fermi level.

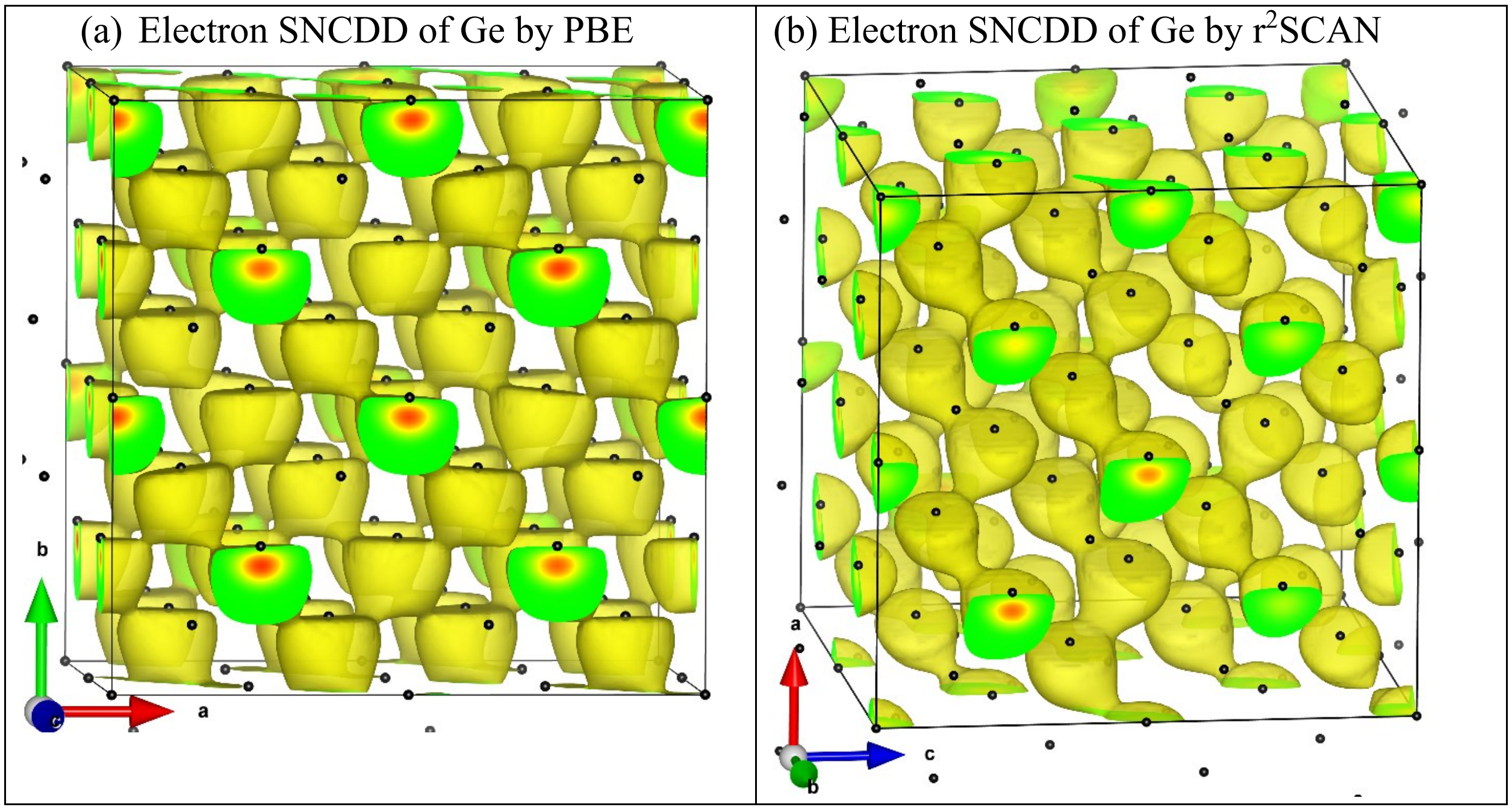

Figure S 17. SNCDD of Ge. PBE predicted electron SNCDD due to charge gain (a, in yellow), and r$^2$SCAN predicted electron SNCDD due to charge gain (b, in yellow). Note that Ge is semiconductor without free electrons and holes at the Fermi level.

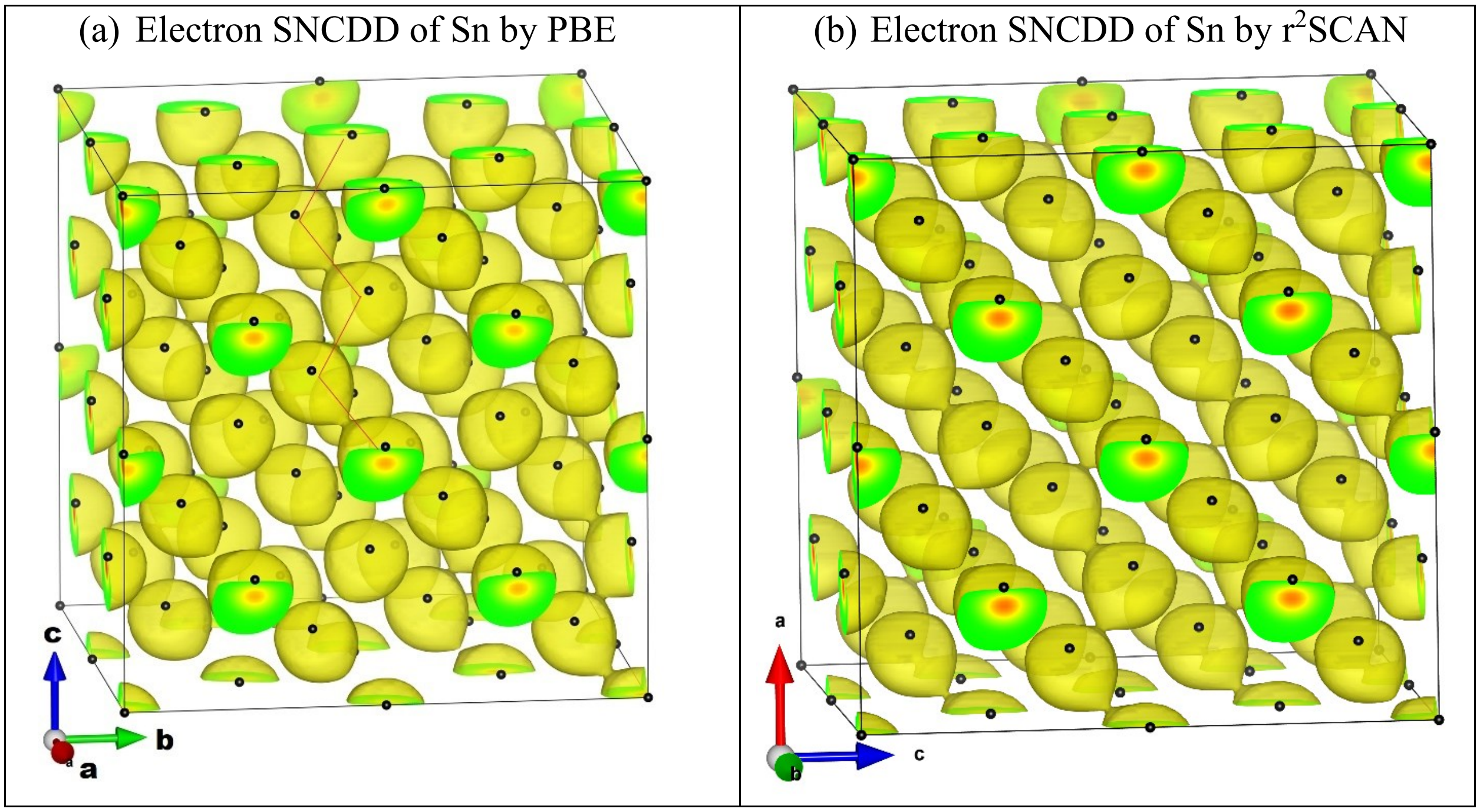

Figure S 18. SNCDD of Sn. PBE predicted electron SNCDD due to charge gain (a, in yellow), and $r^2$SCAN predicted electron SNCDD due to charge gain (b, in yellow). The red lines links one of the 1D-type tunnels formed along [$10\bar{1}$] direction, which is facile for carrier transfer by forming SODTs.

# 8 Supplementary figures of In (A6), Sb (A7), Bi (A7)

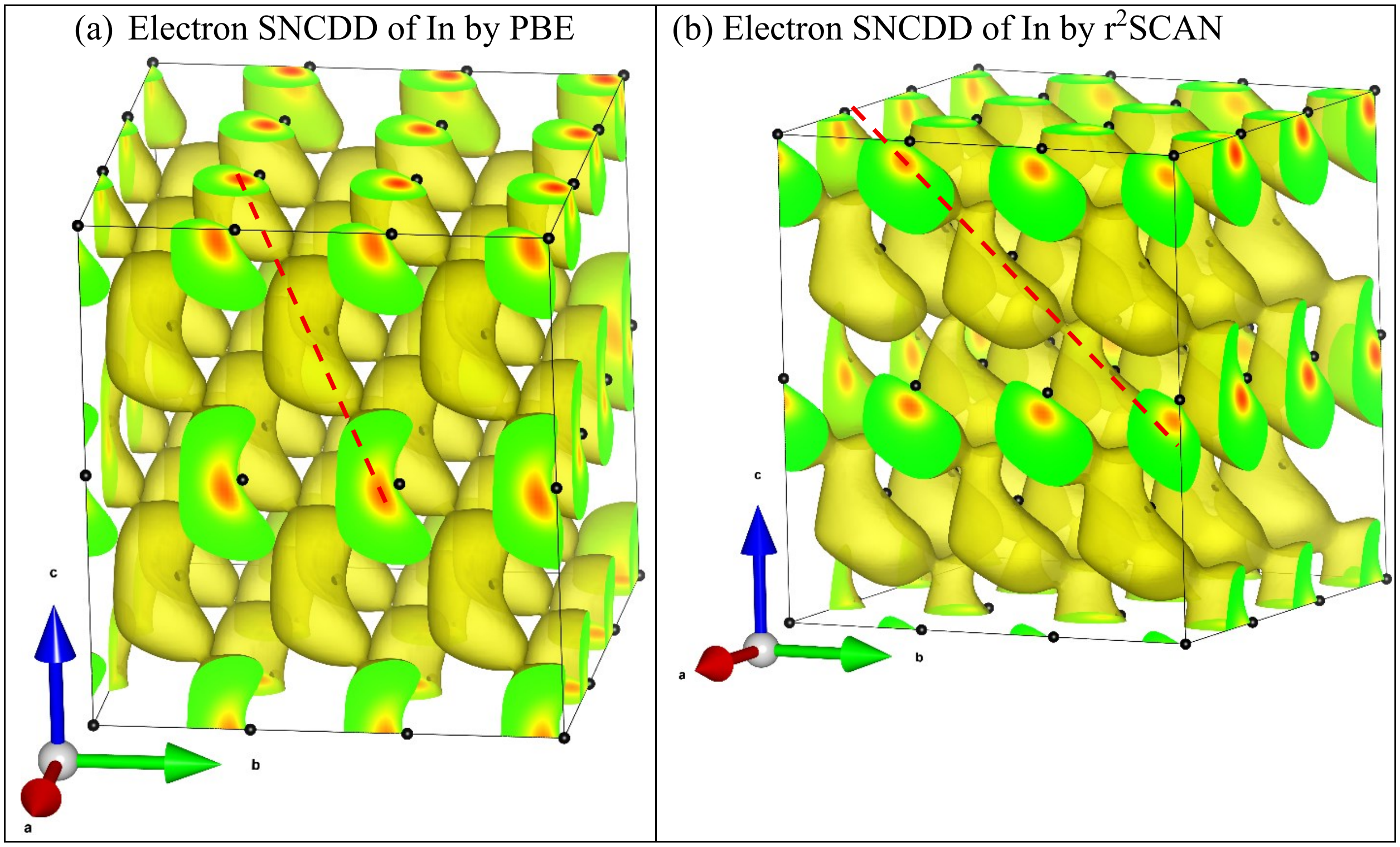

Figure S 19. SNCDD of In. PBE predicted electron SNCDD due to charge gain (a, in yellow), and $r^2$SCAN predicted electron SNCDD due to charge gain (b, in yellow). The presence of SODTs is highlighted using red dashed lines for improved visual clarity.

| (a) Electron SNCDD of Sb by PBE | (b) Electron SNCDD of Sb by r$^2$SCAN |
|---|---|
| 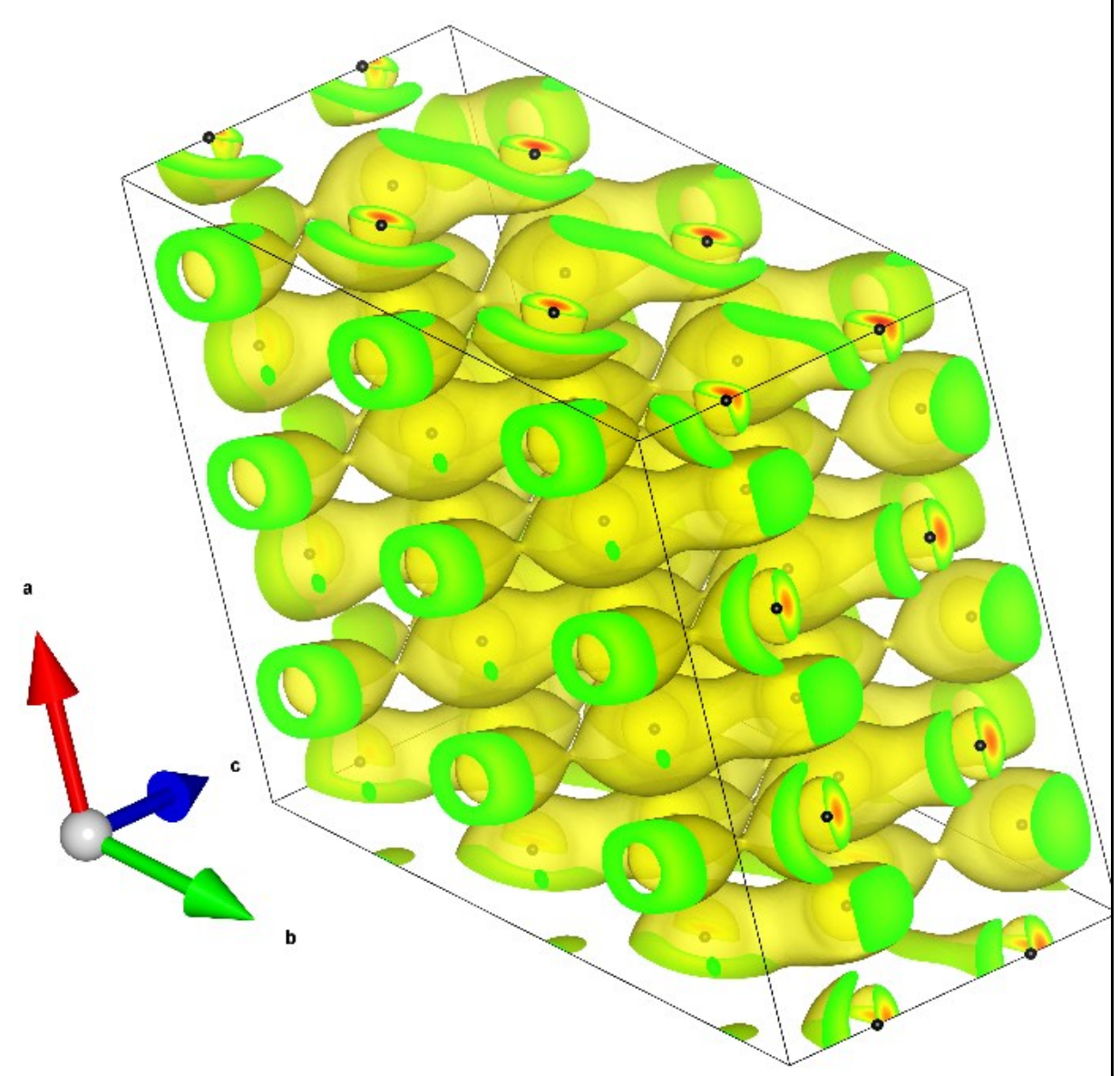 | 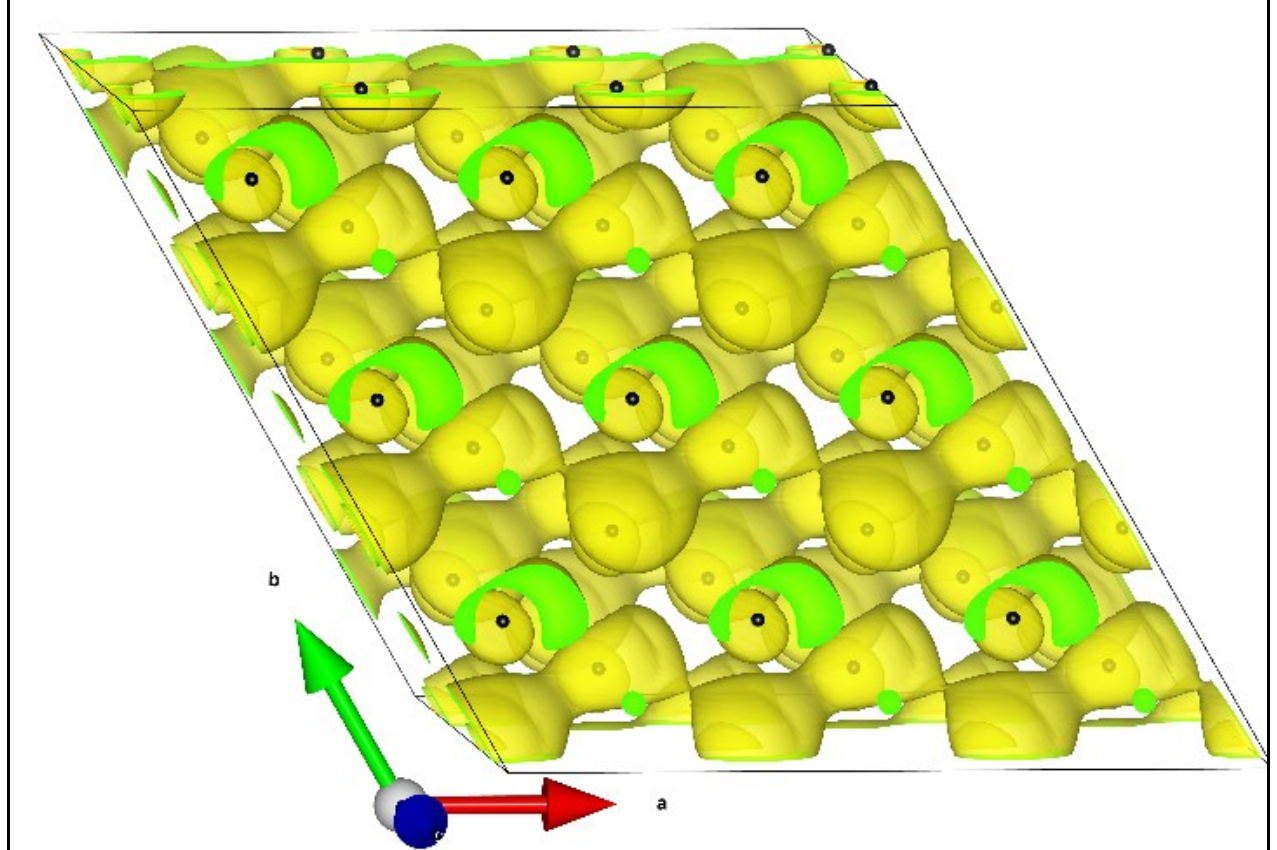 |

Figure S 20. SNCDD of Sb. PBE predicted electron SNCDD due to charge gain (in yellow), and r$^2$SCAN predicted electron SNCDD due to charge gain (b, in yellow), indicating the formation of SODTs along roughly [011] by PBE and [100] by r$^2$SCAN.

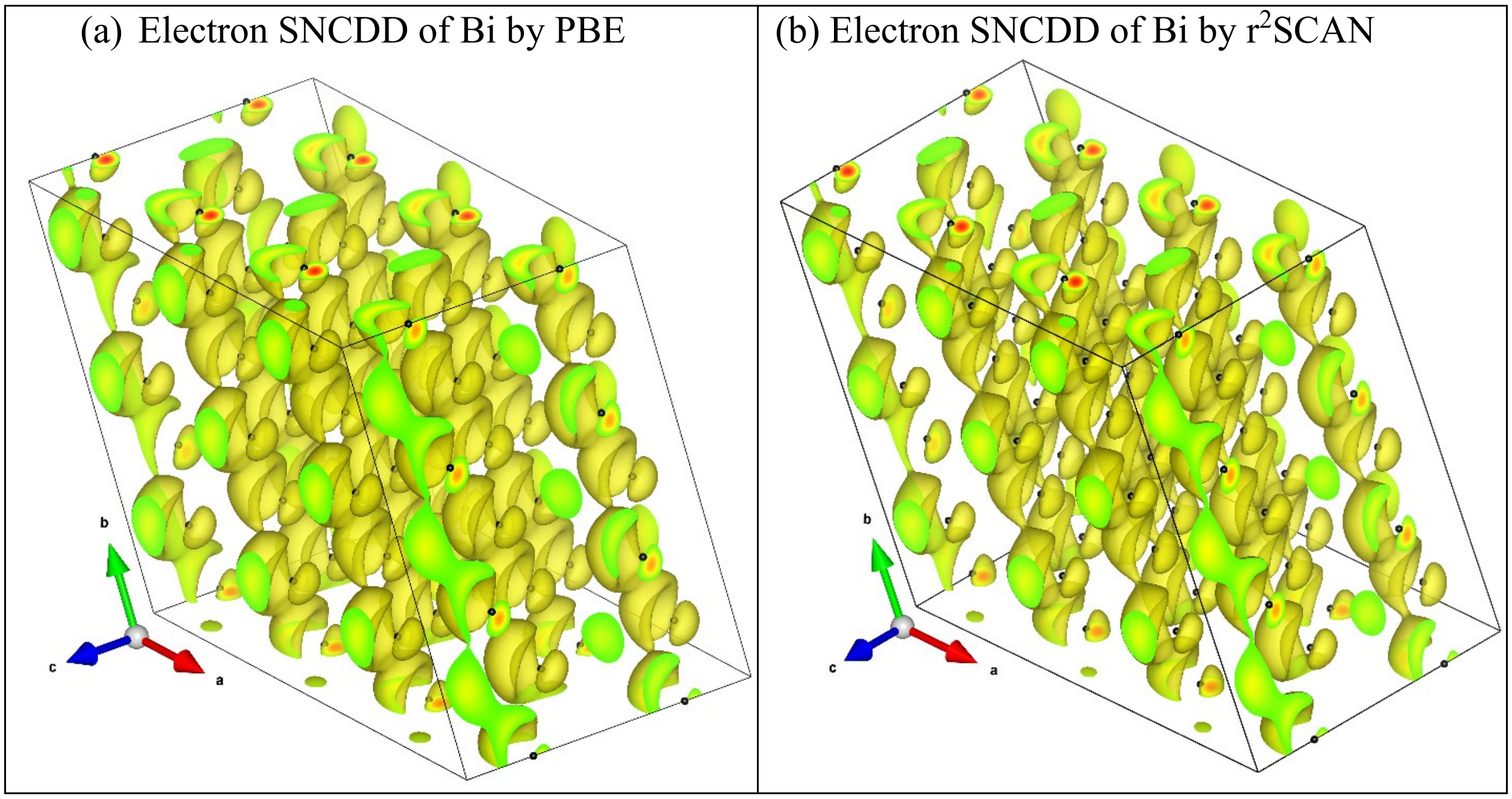

Figure S 21. SNCDD of Bi. PBE predicted electron SNCDD due to charge gain (a, in yellow), and $r^2$SCAN predicted electron SNCDD due to charge gain (b, in yellow), indicating the formation of SODTs along roughly [010] direction.

## 9 Supplementary figures of $YBCO_7$

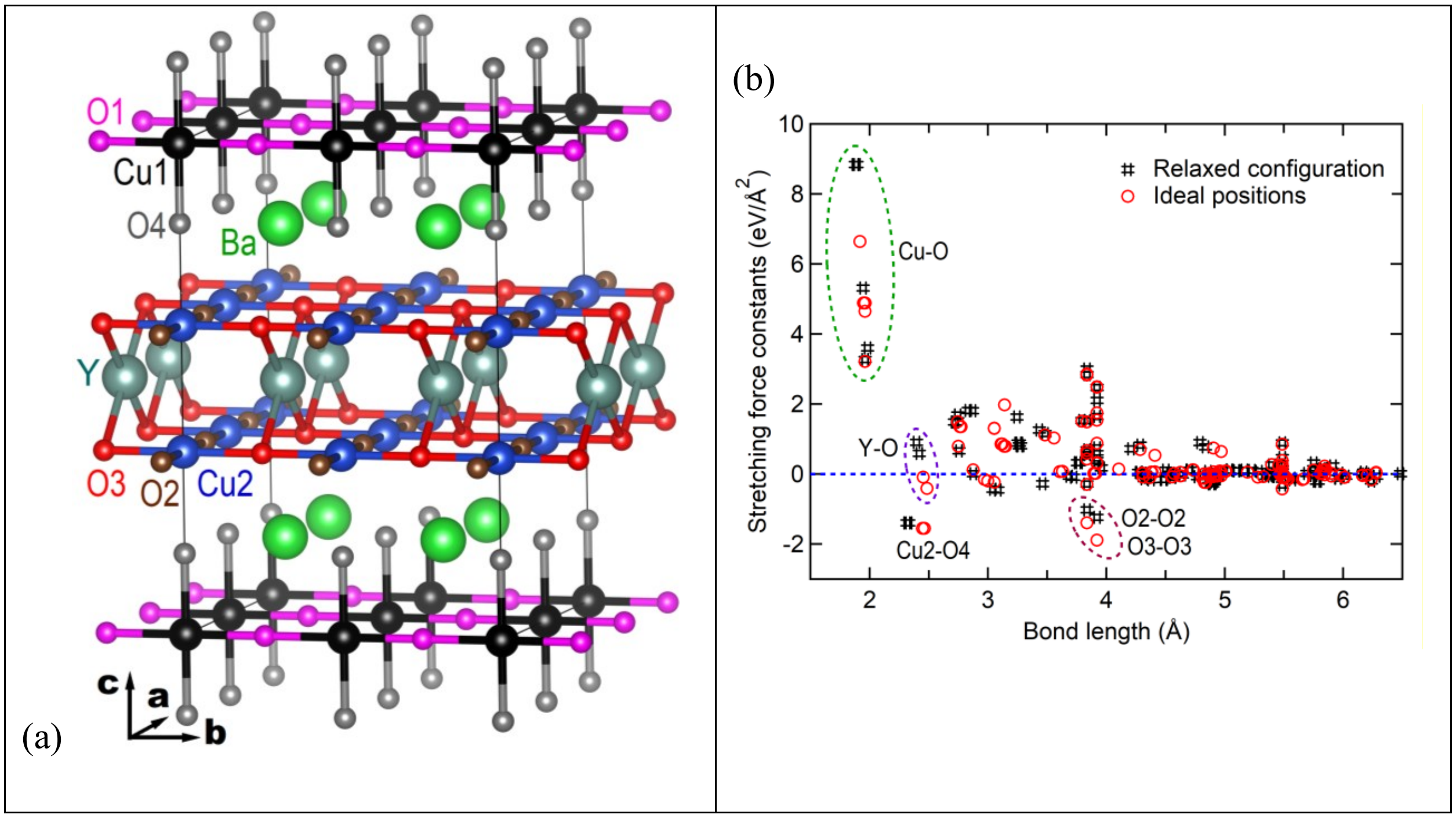

Figure S 22. (a) Undistorted configuration of the 2×2×1 $YBCO_7$ supercell with the bonds connecting key interactions indicated by the stretching force constants (SFCs) from phonon calculations by PBE (b). Crystallographic details of $YBCO_7$ are given in Table S 3, and some key SFCs in red in the undistorted configuration decrease, making the undistorted $YBCO_7$ less stable or even unstable.

## 10 Supplementary figures of YBCO

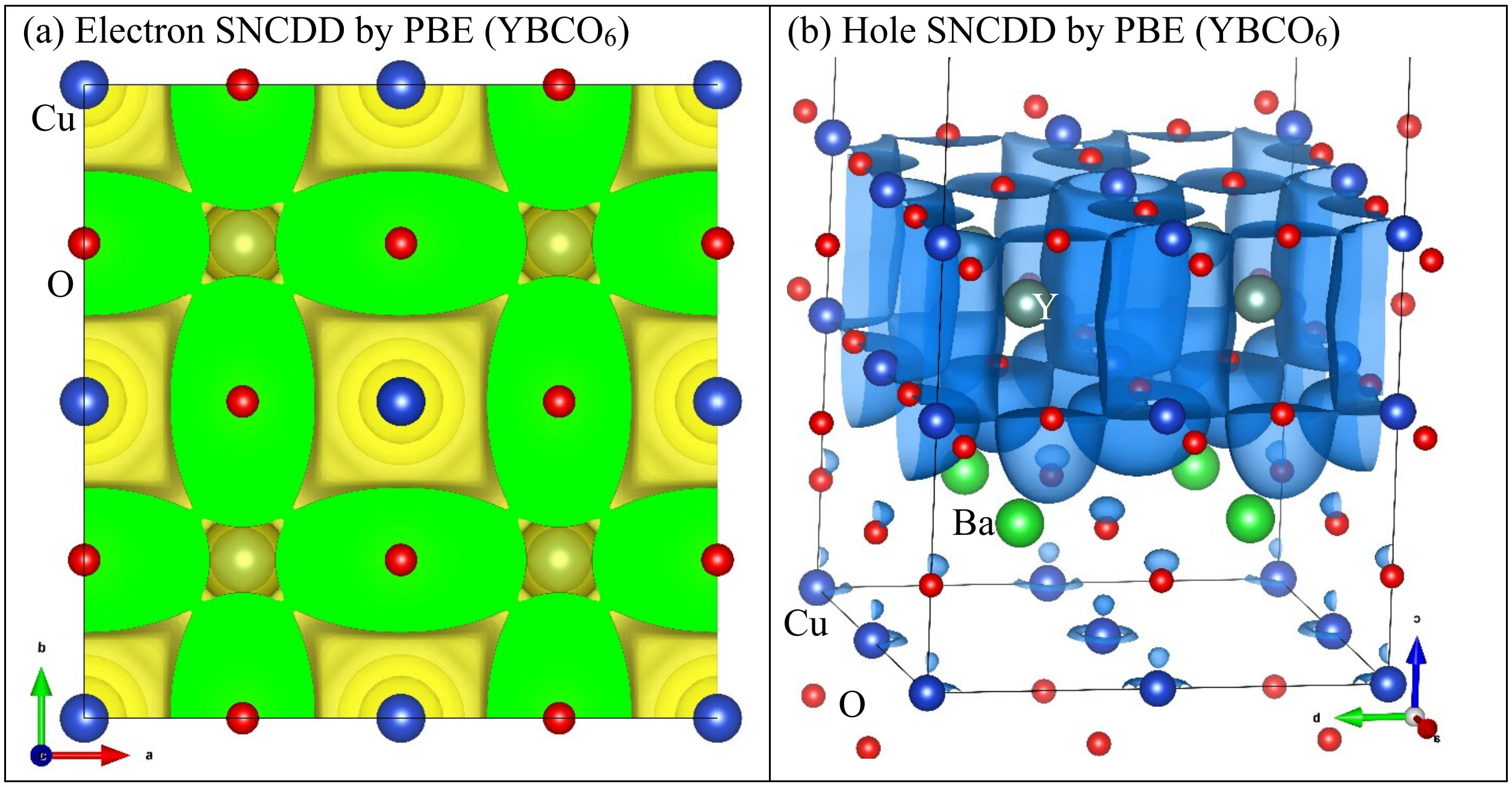

Figure S 23. SNCDD of $YBCO_6$. Partial electron SNCDD (a, in yellow) and hole SNCDD (b, in blue) predicted by PBE, showing the double 2D tunnels formed by the Cu2-O2 atoms (cf., Table S 3) and parallel to the a-b plane for both cases.

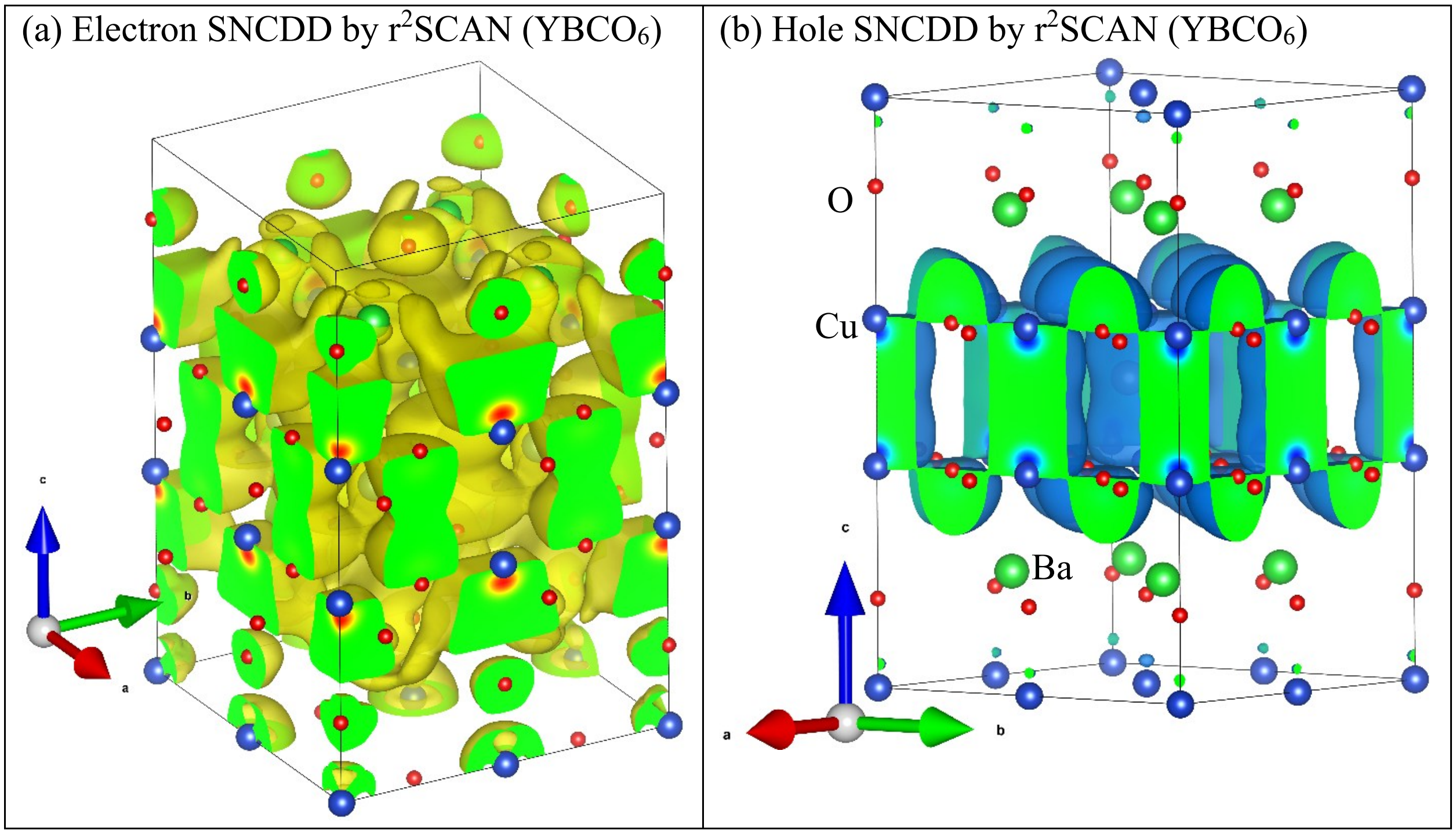

Figure S 24. SNCDD of $YBCO_6$. Electron SNCDD (a, in yellow) and hole SNCDD (b, in blue) predicted by $r^2SCAN$.

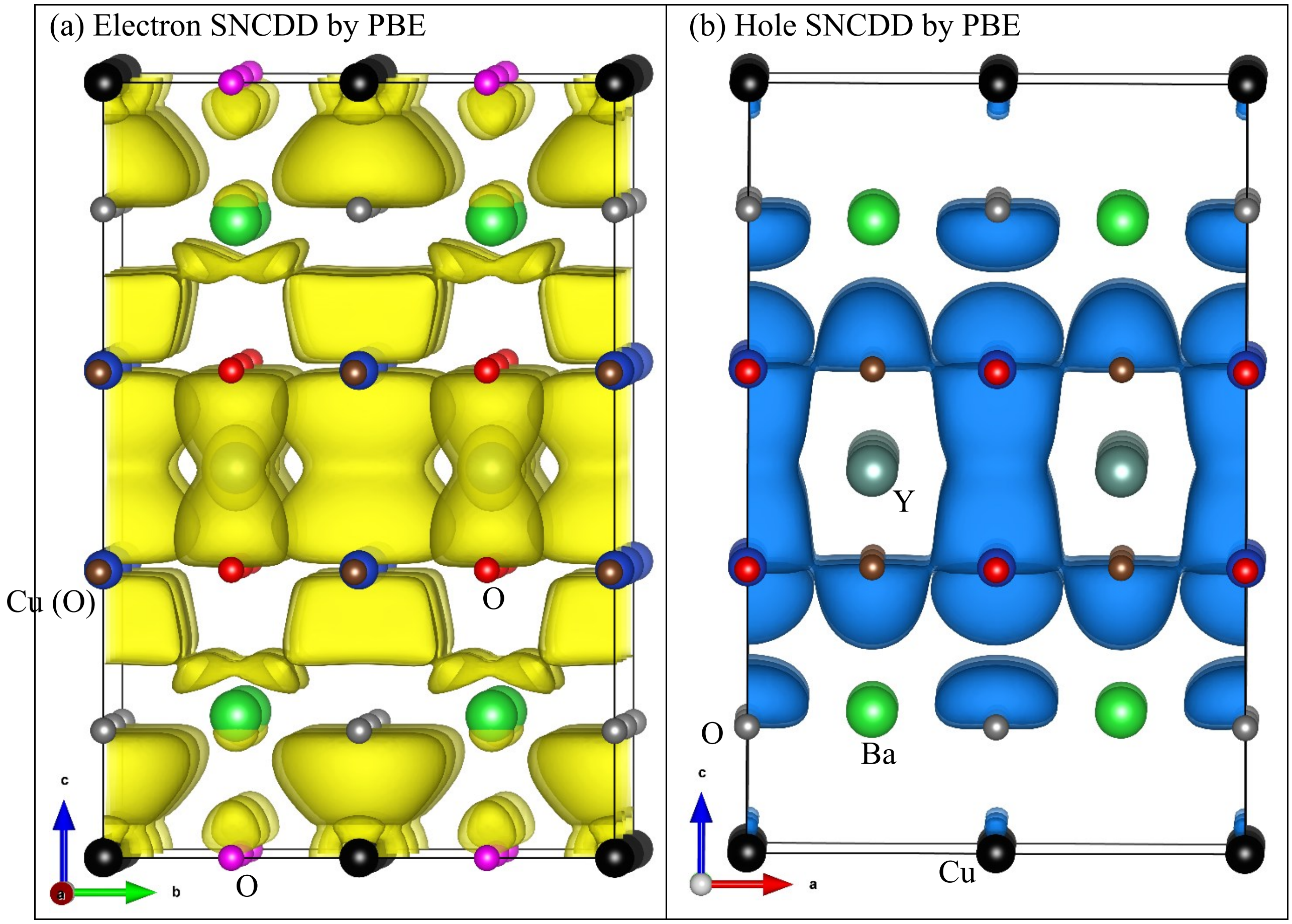

Figure S 25. SNCDD of $YBCO_7$. Predicted electron SNCDD (a, in yellow) and hole SNCDD (b, in blue) by PBE.

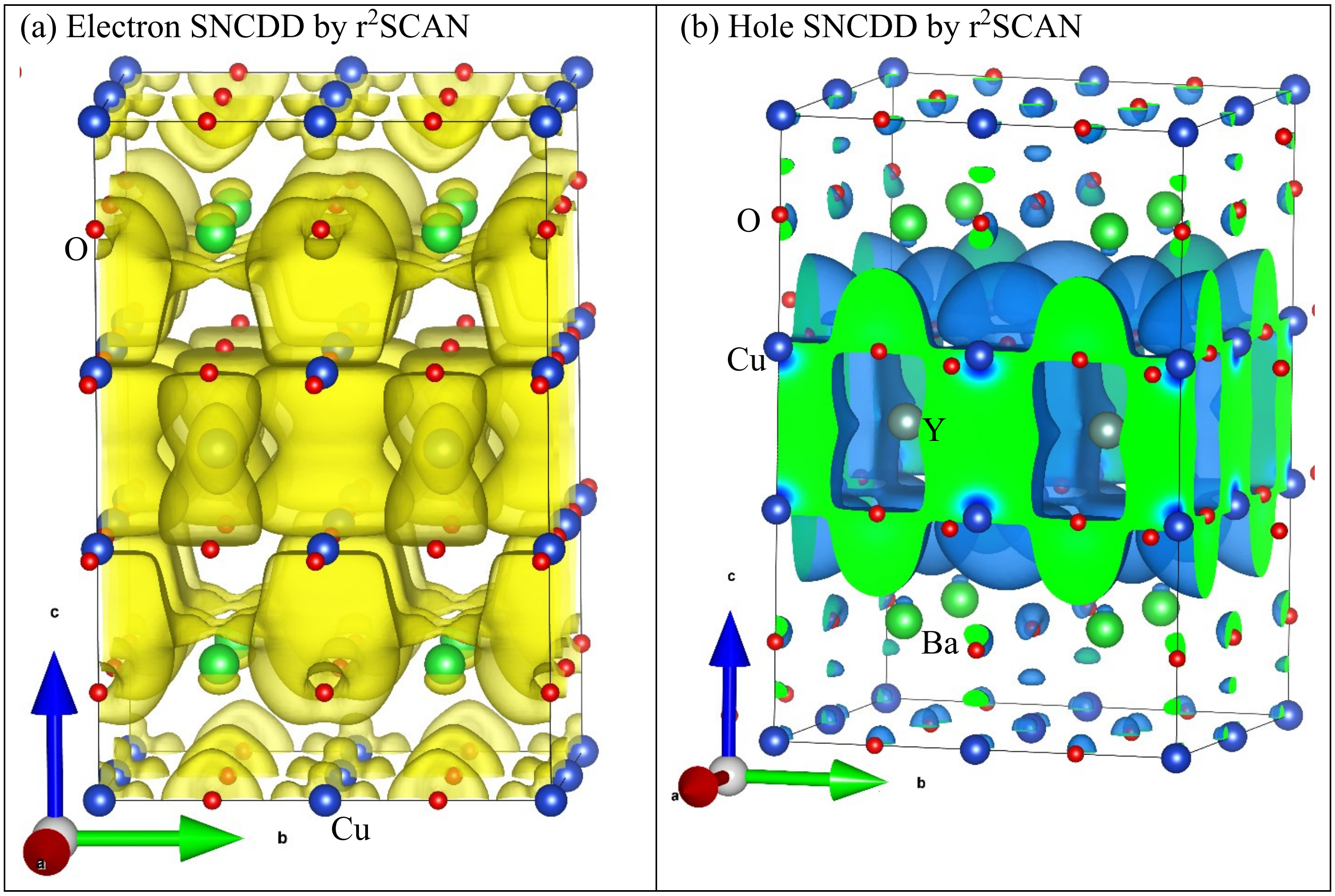

Figure S 26. SNCDD of $YBCO_7$. Predicted electron SNCDD (a, in yellow) and hole SNCDD (b, in blue) by $r^2$SCAN.

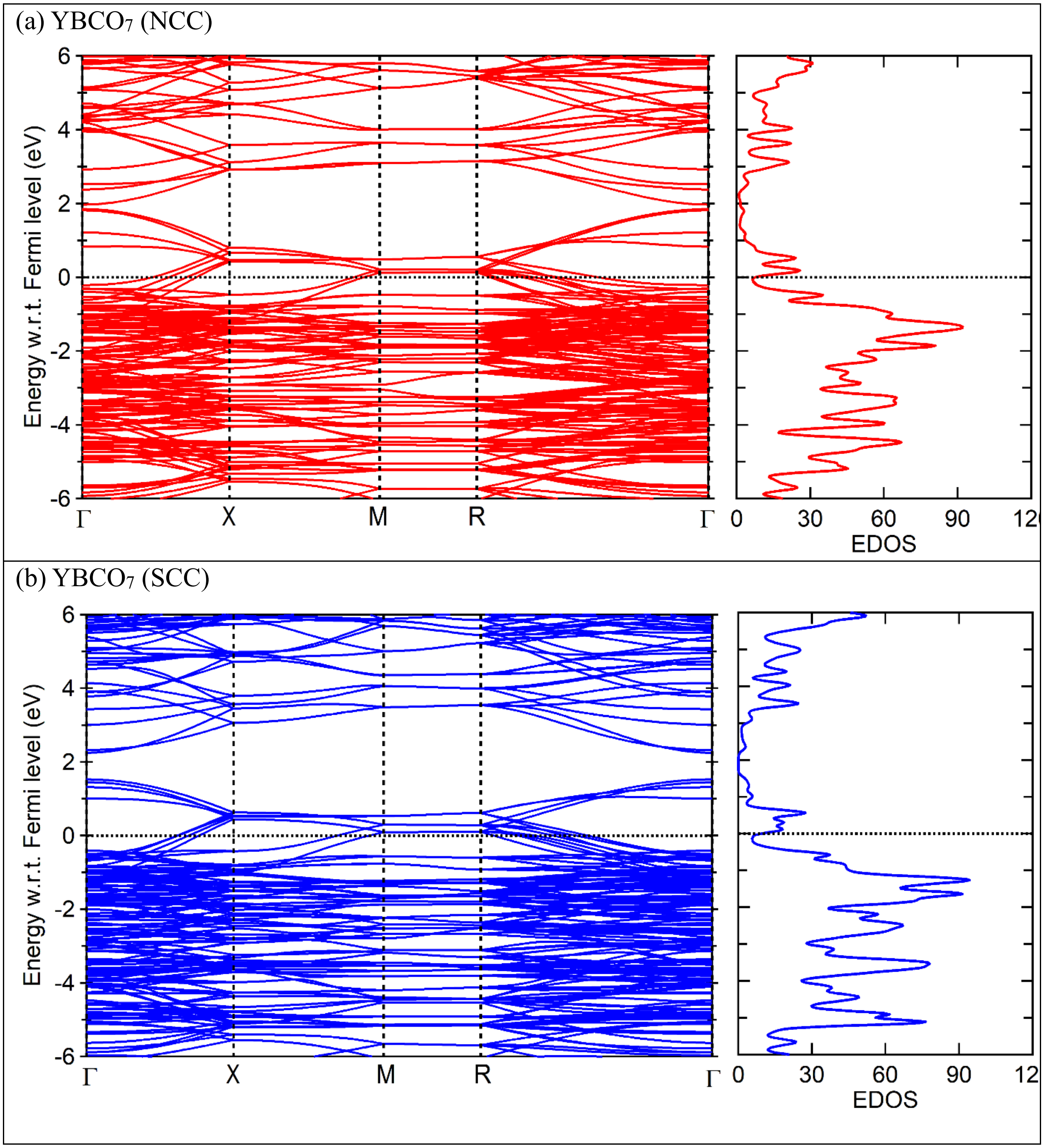

Figure S 27. Band structures and electron density of states (eDOS) of $YBCO_7$ for (a) NCC and (b) SCC by PBE.

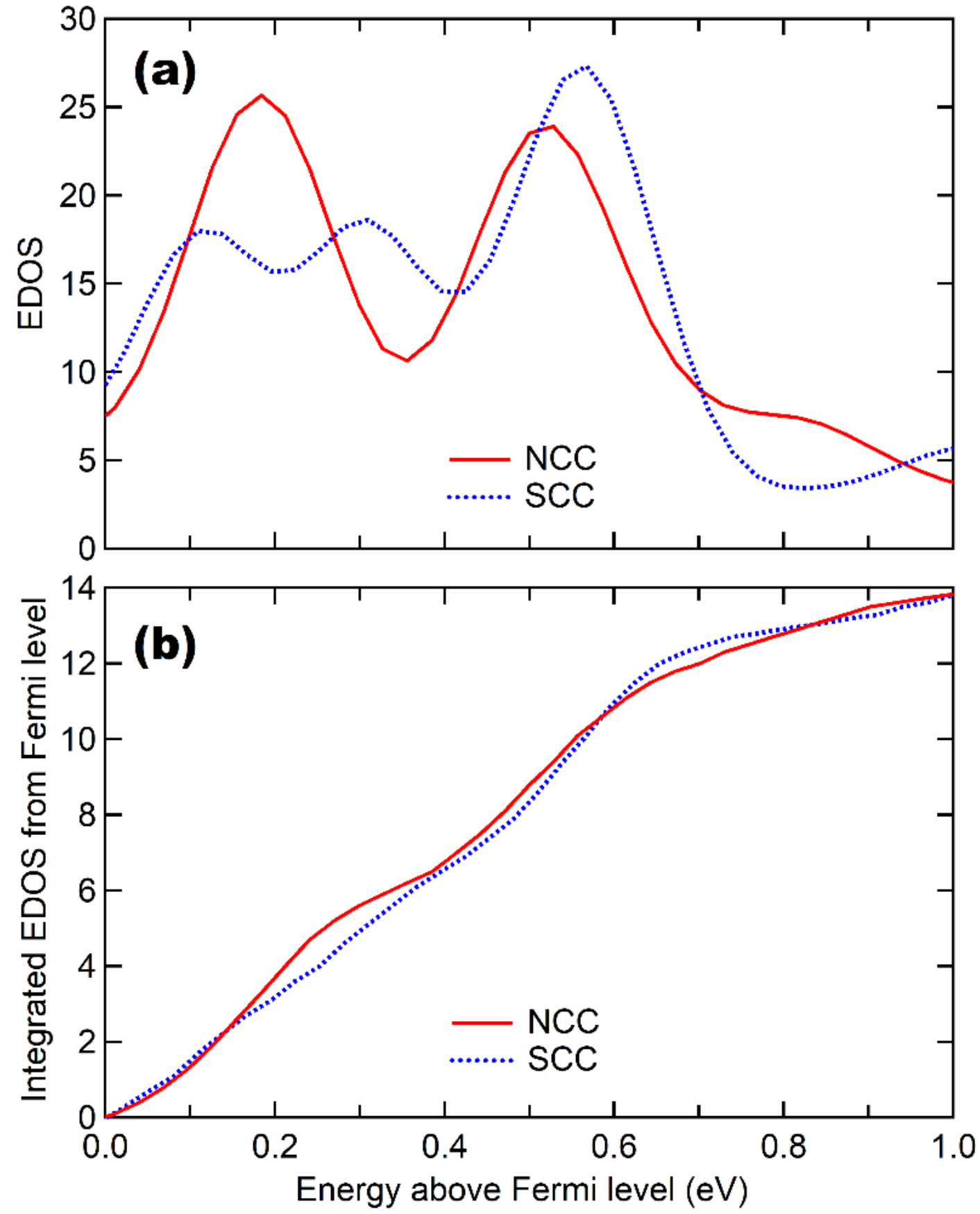

Figure S 28. (a) Zoomed in eDOS and (b) the integrated eDOS above the Fermi level for NCC and SCC of $YBCO_7$ by PBE.

## 11 Supplementary figures of $MgB_2$

(a) Electron SNCDD of $MgB_2$ by $r^2$SCAN

(b) Hole SNCDD of $MgB_2$ by $r^2$SCAN

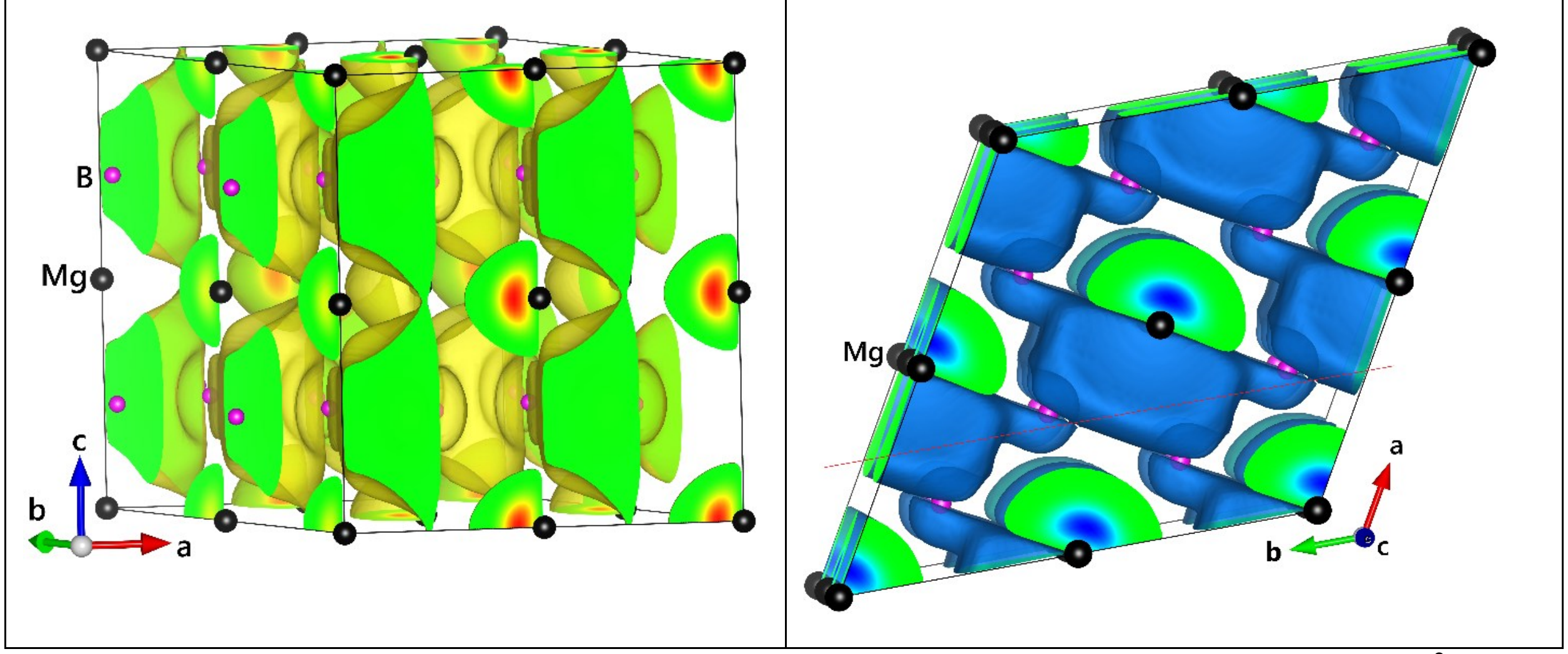

Figure S 29. Electron (a, in yellow) and hole (b, in blue) SNCDDs of $MgB_2$ predicted by $r^2$SCAN, showing SODTs along c-axis for electron and b-axis for hole.